\documentclass[aps,preprint]{revtex4}%
\usepackage{amsfonts}
\usepackage{amsmath}
\usepackage{amssymb}
\usepackage{subfigure}
\usepackage{graphicx}
\usepackage{array}%
\begin{document}
\preprint{CTP-SCU/2018001}
\title{Holographic DC Conductivity for Backreacted Nonlinear Electrodynamics with
Momentum Dissipation }
\author{Peng Wang}
\email{pengw@scu.edu.cn}
\author{Houwen Wu}
\email{wuhouwen@stu.scu.edu.cn}
\author{Haitang Yang}
\email{hyanga@scu.edu.cn}
\affiliation{Center for Theoretical Physics, College of Physical Science and Technology,
Sichuan University, Chengdu, 610064, PR China}

\begin{abstract}
We consider a holographic model with the charge current dual to a general
nonlinear electrodynamics (NLED) field. Taking into account the backreaction
of the NLED field on the geometry and introducing axionic scalars to generate
momentum dissipation, we obtain expressions for DC conductivities with a
finite magnetic field. The properties of the in-plane resistance are examined
in several NLED models. For Maxwell-Chern-Simons electrodynamics, negative
magneto-resistance and Mott-like behavior could appear in some parameter space
region. Depending on the sign of the parameters, we expect the NLED models to
mimic some type of weak or strong interactions between electrons. In the
latter case, negative magneto-resistance and Mott-like behavior can be
realized at low temperatures. Moreover, the Mott insulator to metal transition
induced by a magnetic field is also observed at low temperatures.

\end{abstract}
\keywords{}\maketitle
\tableofcontents

\bigskip



\section{Introduction}

Gauge/gravity duality
\cite{IN-Banks:1996vh,IN-Maldacena:1997re,IN-Witten:1998qj} has provided
powerful tools for exploring the behavior of strongly coupled quantum phases
of matter, and some remarkable progresses have been made
\cite{IN-Gubser:2008px,IN-Hartnoll:2008vx,IN-Lee:2008xf,IN-Liu:2009dm,IN-Cubrovic:2009ye}%
. Conductivity is an important transport quantity in condensed matter, and the
gauge/gravity duality provides a framework to compute it for strongly
interacting field theories.

Studying the behavior of the conductivity in the presence of external magnetic
fields can help us to better understand the transport properties of materials.
For normal metals, the resistance is a monotonically increasing function of
the magnetic field \cite{IN-Wannier1972}, which appears as positive
magneto-resistance. However, negative magneto-resistance has been observed in
several experiments \cite{IN-Negishi,IN-Li,IN-Kim}. On the other hand, the
behavior of negative magneto-resistance was found in strongly coupled
holographic chiral anomalous systems
\cite{IN-Jimenez-Alba:2014iia,IN-Jimenez-Alba:2015awa,IN-Landsteiner:2014vua,IN-Sun:2016gpy}%
. In \cite{IN-Baumgartner:2017kme}, it showed that negative magneto-resistance
could also arise in nonanomalous relativistic fluids due to the distinctive
gradient expansion. Note that the transport phenomena in the presence of Weyl
corrections have also been discussed in
\cite{IN-Mokhtari:2017vyz,IN-Chu:2018ksb,Chu:2018ntx}. Recently, the
magnetotransport of a strongly interacting system in $2+1$ dimensions was
examined in a holographic Dirac-Born-Infeld model in
\cite{IN-Kiritsis:2016cpm,IN-Cremonini:2017qwq}. Negative magnetoresistance
was found for a family of dynoic solutions in \cite{IN-Cremonini:2017qwq}. The
DC conductivity in the probe DBI case with the vanishing magnetic field was
also discussed in \cite{IN-Charmousis:2010zz}.

Mott insulators can be parent materials of high $T_{c}$ cuprate
superconductors. A Mott insulator has an insulating ground state driven by
Coulomb repulsion. Mott-like behavior is that strong interactions between
electrons would prevent the charge carriers to efficiently transport charges.
Constructing a holographic model describing Mott insulators is still a
challenging task. In
\cite{IN-Edalati:2010ww,IN-Edalati:2010ge,IN-Wu:2012fk,IN-Ling:2014bda},
dynamically generating a Mott gap has been proposed in holographic models by
considering fermions with dipole coupling. A holographic construction of the
large-$N$ Bose-Hubbard model was presented in \cite{IN-Fujita:2014mqa}, and
the model admitted Mott insulator ground states in the limit of large Coulomb
repulsion. Some other holographic models dual to Mott insulators include
\cite{IN-Ling:2015epa,IN-Nishioka:2009zj,IN-Kiritsis:2015oxa}. Recently, a
holographic model using a particular type of NLED, namely iDBI, was proposed
in \cite{IN-Baggioli:2016oju} to mimic interactions between electrons by
self-interactions of the NLED field. It showed that Mott-like behavior
appeared for large enough self-interaction strength.

In this paper, we extend the analysis of the magnetotransport in a holographic
Dirac-Born-Infeld model in \cite{IN-Cremonini:2017qwq} to a general NLED
model. As in \cite{IN-Cremonini:2017qwq}, our analysis is performed in a full
backreacted fashion. To break translational symmetry, we follow the method in
\cite{IN-Andrade:2013gsa} to add axionic scalars, which depend on the spatial
directions linearly.

The rest of this article is organized as follows. In section \ref{Sec:HS}, we
set up our holographic model. The expressions for the DC conductivities with a
finite magnetic field are obtained in section \ref{Sec:DCC}. Some limiting
cases, including high temperature limit, are then discussed. In section
\ref{Sec:Examples}, the dependence of the in-plane resistance on the
temperature, the charge density and the magnetic field are investigated for
Maxwell, Maxwell-Chern-Simons, Born-Infeld, square and\ logarithmic
electrodynamics. In section \ref{Sec:Con}, we summarize our results and
conclude with a brief discussion.

\section{Holographic Setup}

\label{Sec:HS}

Consider a 4-dimensional model of gravity coupled to a nonlinear
electromagnetic field $A_{a}$ and two axions $\psi_{I}$ with action given by%
\begin{equation}
S=\int d^{4}x\sqrt{-g}\left[  R-2\Lambda-\frac{1}{2}\sum_{I=1}^{2}\left(
\partial\psi_{I}\right)  ^{2}+\mathcal{L}\left(  s,p\right)  \right]  ,
\label{eq:Action}%
\end{equation}
where $\Lambda=-\frac{3}{l^{2}}$, and we take $16\pi G=1$ for simplicity. In
the action $\left(  \ref{eq:Action}\right)  $, we assume that the generic NLED
Lagrangian is $\mathcal{L}\left(  s,p\right)  $, where we build two
independent nontrivial scalars using $F_{ab}=\partial_{a}A_{b}-\partial
_{b}A_{a}$ and none of its derivatives:
\begin{equation}
s=-\frac{1}{4}F^{ab}F_{ab}\text{ and }p=-\frac{1}{8}\epsilon^{abcd}%
F_{ab}F_{cd}\text{;}%
\end{equation}
$\epsilon^{abcd}\equiv-\left[  a\text{ }b\text{ }c\text{ }d\right]  /\sqrt
{-g}$ is a totally antisymmetric Lorentz tensor, and $\left[  a\text{ }b\text{
}c\text{ }d\right]  $ is the permutation symbol. We also assume that the NLED
Lagrangian would reduce to the form of Maxwell-Chern-Simons Lagrangian for
small fields:%
\begin{equation}
\mathcal{L}\left(  s,p\right)  \approx s+\theta p,
\end{equation}
where, for later convenience, we define $\theta\equiv\mathcal{L}^{\left(
0,1\right)  }\left(  0,0\right)  $. Note that we set the AdS radius $l=1$ hereafter.

Varying the action $\left(  \ref{eq:Action}\right)  $ with respect to $g_{ab}%
$, $A_{a}$, and $\psi_{I}$, we find that the equations of motion are%
\begin{align}
R_{ab}-\frac{1}{2}Rg_{ab}-\frac{3}{l^{2}}g_{ab}  &  =\frac{T_{ab}}{2}%
\text{,}\nonumber\\
\nabla_{a}G^{ab}  &  =0\text{,}\\
\partial_{\mu}\left(  \sqrt{-g}\partial^{\mu}\psi_{I}\right)   &
=0\text{,}\nonumber
\end{align}
where $T_{ab}$ is the energy-momentum tensor:%
\begin{equation}
T_{ab}=g_{ab}\left[  -\frac{1}{2}\sum_{I=1}^{2}\left(  \partial\psi
_{I}\right)  ^{2}+\mathcal{L}\left(  s,p\right)  -p\frac{\partial
\mathcal{L}\left(  s,p\right)  }{\partial p}\right]  +\sum_{I=1}^{2}\left(
\partial_{a}\psi_{I}\right)  \left(  \partial_{b}\psi_{I}\right)
+\frac{\partial\mathcal{L}\left(  s,p\right)  }{\partial s}F_{a}^{\text{ }%
c}F_{bc}\text{,}%
\end{equation}
and we define%
\begin{equation}
G^{ab}=-\frac{\partial\mathcal{L}\left(  s,p\right)  }{\partial F_{ab}}%
=\frac{\partial\mathcal{L}\left(  s,p\right)  }{\partial s}F^{ab}+\frac{1}%
{2}\frac{\partial\mathcal{L}\left(  s,p\right)  }{\partial p}\epsilon
^{abcd}F_{cd}. \label{eq:Gab}%
\end{equation}

To construct a black brane solution with asymptotic AdS spacetime, we take the
following ansatz for the metric, the NLED field and the axions%
\begin{align}
ds^{2}  &  =-f\left(  r\right)  dt^{2}+\frac{dr^{2}}{f\left(  r\right)
}+r^{2}\left(  dx^{2}+dy^{2}\right)  \text{,}\nonumber\\
A  &  =A_{t}\left(  r\right)  dt+\frac{h}{2}\left(  xdy-ydx\right)
\text{,}\label{eq:BBAnsatz}\\
\psi_{1}  &  =\alpha x\text{, and }\psi_{2}=\alpha y\text{,}\nonumber
\end{align}
where $h$ denotes the magnitude of the magnetic field. The axions are
responsible for the breaking the translational invariance and generating
momentum dissipation. The equations of motion then take the form:%
\begin{align}
f\left(  r\right)  -3r^{2}+rf^{\prime}\left(  r\right)   &  =-\frac{\alpha
^{2}}{2}+\frac{r^{2}}{2}\left[  \mathcal{L}\left(  s,p\right)  +A_{t}^{\prime
}\left(  r\right)  G^{rt}\right]  ,\label{eq:ttEOM}\\
2f^{\prime}\left(  r\right)  -6r+rf^{\prime\prime}\left(  r\right)   &
=r\left[  \mathcal{L}\left(  s,p\right)  +hG^{xy}\right]  ,\label{eq:rrEOM}\\
\left[  r^{2}G^{rt}\right]  ^{\prime}  &  =0\text{.} \label{eq:NLEDEOM}%
\end{align}
It can be shown that eqns. $\left(  \ref{eq:ttEOM}\right)  $ and $\left(
\ref{eq:NLEDEOM}\right)  $ guarantee that eqn. $\left(  \ref{eq:rrEOM}\right)
$ always holds. Eqn. $\left(  \ref{eq:NLEDEOM}\right)  $ leads to%
\begin{equation}
G^{tr}=\frac{\rho}{r^{2}}\text{,} \label{eq:Grt}%
\end{equation}
where $\rho$ is a constant. One has $f\left(  r_{h}\right)  =0$ at the horizon
$r=r_{h}$, and the Hawking temperature of the black brane is given by%
\begin{equation}
T=\frac{f^{\prime}\left(  r_{h}\right)  }{4\pi}\text{.}%
\end{equation}
Hence at $r=r_{h}$, eqn. $\left(  \ref{eq:ttEOM}\right)  $ reduces to%
\begin{equation}
-3r_{h}^{2}+4\pi r_{h}T=-\frac{\alpha^{2}}{2}+\frac{r_{h}^{2}}{2}\left[
\mathcal{L}\left(  s_{h},p_{h}\right)  +A_{t}^{\prime}\left(  r_{h}\right)
G_{h}^{rt}\right]  , \label{eq:HT}%
\end{equation}
where%
\begin{align}
s_{h}  &  =\frac{A_{t}^{\prime2}\left(  r_{h}\right)  }{2}-\frac{h^{2}}%
{2r_{h}^{4}}\text{,}\nonumber\\
\text{ }p_{h}  &  =-\frac{hA_{t}^{\prime}\left(  r_{h}\right)  }{r_{h}^{2}%
}\text{,}\label{eq:shph}\\
G_{h}^{rt}  &  =-\mathcal{L}^{\left(  1,0\right)  }\left(  s_{h},p_{h}\right)
A_{t}^{\prime}\left(  r_{h}\right)  +\mathcal{L}^{\left(  0,1\right)  }\left(
s_{h},p_{h}\right)  \frac{h}{r_{h}^{2}}\text{.}\nonumber
\end{align}

\section{DC Conductivity}

\label{Sec:DCC}

Via gauge/gravity duality, the black brane solution $\left(  \ref{eq:BBAnsatz}%
\right)  $ describes an equilibrium state at finite temperature $T$, which is
given by eqn. $\left(  \ref{eq:HT}\right)  $. The NLED field is a U$\left(
1\right)  $ gauge field and dual to a conserved current $\mathcal{J}^{\mu}$ in
the boundary theory. In this section, we calculate the DC conductivities for
$\mathcal{J}^{\mu}$ using the method developed in
\cite{In-Donos:2014uba,IN-Blake:2014yla}.

\subsection{Derivation of DC Conductivity}

To calculate the DC conductivities, we consider the perturbations of the form:%
\begin{equation}
\delta g_{ti}=r^{2}h_{ti}\left(  r\right)  \text{, }\delta g_{ri}=r^{2}%
h_{ri}\left(  r\right)  \text{, }\delta A_{i}=-E_{i}t+a_{i}\left(  r\right)
\text{, }\delta\psi_{I}=\chi_{I}\left(  r\right)  \text{,} \label{eq:purb}%
\end{equation}
where $i=x$, $y$ and $I=1,2$. The fields $a_{i}\left(  r\right)  $ do not
appear explicitly in the NLED Lagrangian $\mathcal{L}\left(  s,p\right)  $.
Thus, the conjugate momentum of the field $a_{i}\left(  r\right)  $ with
respect to $r$-foliation is radially independent:%
\begin{equation}
\partial_{r}\Pi^{i}=0\text{,}%
\end{equation}
where the conjugate current is%
\begin{equation}
\Pi^{i}=\frac{\partial\mathcal{L}\left(  s,p\right)  }{\partial\left(
a_{i}^{\prime}\left(  r\right)  \right)  }=\frac{\partial\mathcal{L}\left(
s,p\right)  }{\partial\left(  \partial_{r}A_{i}\right)  }=\sqrt{-g}G^{ir}.
\end{equation}
Similarly, the conjugate momentum of the field $A_{t}\left(  r\right)  $ is
also a constant flux%
\begin{equation}
\partial_{r}\Pi^{t}=0\text{,}%
\end{equation}
where one has%
\begin{equation}
\Pi^{t}=\frac{\partial\mathcal{L}\left(  s,p\right)  }{\partial\left(
A_{t}^{\prime}\left(  r\right)  \right)  }. \label{eq:Pit}%
\end{equation}
We can then compute the expectation value of $\mathcal{J}^{t}$ for the
boundary theory by%
\begin{equation}
\left\langle \mathcal{J}^{t}\right\rangle =\Pi^{t}.
\end{equation}
Using eqns. $\left(  \ref{eq:Gab}\right)  $, $\left(  \ref{eq:BBAnsatz}%
\right)  $, $\left(  \ref{eq:Grt}\right)  $, $\left(  \ref{eq:purb}\right)  $,
and $\left(  \ref{eq:Pit}\right)  $, we find that, at the linearized order,
\begin{equation}
\left\langle \mathcal{J}^{t}\right\rangle =\rho\text{,}%
\end{equation}
which means that $\rho$ can be interpreted as the charge density in the dual
field theory. Evaluating eqn. $\left(  \ref{eq:Pit}\right)  $ at $r=r_{h}$
gives%
\begin{equation}
\rho=\mathcal{L}^{\left(  1,0\right)  }\left(  s_{h},p_{h}\right)
A_{t}^{\prime}\left(  r_{h}\right)  -\mathcal{L}^{\left(  0,1\right)  }\left(
s_{h},p_{h}\right)  \frac{h}{r_{h}^{2}}. \label{eq:rho}%
\end{equation}
The charge currents in the dual theory are given by%
\begin{equation}
\left\langle \mathcal{J}^{i}\right\rangle =\Pi^{i}\text{,}%
\end{equation}
which lead to
\begin{align}
\left\langle \mathcal{J}^{x}\right\rangle  &  =-\mathcal{L}^{\left(
1,0\right)  }\left(  s,p\right)  \left[  f\left(  r\right)  a_{x}^{\prime
}\left(  r\right)  +hf\left(  r\right)  h_{ry}\left(  r\right)  +r^{2}%
A_{t}^{\prime}\left(  r\right)  h_{tx}\left(  r\right)  \right]
-\mathcal{L}^{\left(  0,1\right)  }\left(  s,p\right)  E_{y}\text{,}%
\nonumber\\
\left\langle \mathcal{J}^{y}\right\rangle  &  =-\mathcal{L}^{\left(
1,0\right)  }\left(  s,p\right)  \left[  f\left(  r\right)  a_{y}^{\prime
}\left(  r\right)  -hf\left(  r\right)  h_{rx}\left(  r\right)  +r^{2}%
A_{t}^{\prime}\left(  r\right)  h_{ty}\left(  r\right)  \right]
+\mathcal{L}^{\left(  0,1\right)  }\left(  s,p\right)  E_{x}\text{.}
\label{eq:Jxy}%
\end{align}

To express $\left\langle \mathcal{J}^{i}\right\rangle $ in terms of $E_{i}$,
we first consider the constraints of regularity on the metric and fields
around the horizon \cite{IN-Blake:2014yla}:
\begin{align}
f\left(  r\right)   &  =4\pi T\left(  r-r_{h}\right)  +\cdots\text{,}%
\nonumber\\
A_{t}\left(  r\right)   &  =A_{t}^{\prime}\left(  r_{h}\right)  \left(
r-r_{h}\right)  +\cdots\text{,}\nonumber\\
a_{i}\left(  r\right)   &  =-\frac{E_{i}}{4\pi T}\ln\left(  r-r_{h}\right)
+\cdots\text{,}\label{eq:reg}\\
h_{ri}\left(  r\right)   &  =\frac{h_{ti}\left(  r\right)  }{f\left(
r\right)  }+\cdots\text{,}\nonumber\\
\chi_{I}\left(  r\right)   &  =\chi_{I}\left(  r_{h}\right)  +\cdots
\text{.}\nonumber
\end{align}
We then consider the $tx$ and $ty$ component of the perturbed Einstein's
equations:%
\begin{align}
h_{tx}\left(  r\right)  \left[  -\frac{\alpha^{2}}{\mathcal{L}^{\left(
1,0\right)  }\left(  s,p\right)  }-\frac{h^{2}}{r^{2}}\right]  -\frac{hE_{y}%
}{r^{2}}+A_{t}^{\prime}\left(  r\right)  f\left(  r\right)  \left[
a_{x}^{\prime}\left(  r\right)  +hh_{ry}\left(  r\right)  \right]   &
=0,\nonumber\\
h_{ty}\left(  r\right)  \left[  -\frac{\alpha^{2}}{\mathcal{L}^{\left(
1,0\right)  }\left(  s,p\right)  }-\frac{h^{2}}{r^{2}}\right]  +\frac{hE_{x}%
}{r^{2}}+A_{t}^{\prime}\left(  r\right)  f\left(  r\right)  \left[
a_{y}^{\prime}\left(  r\right)  -hh_{rx}\left(  r\right)  \right]   &
=0\text{.} \label{eq:txty}%
\end{align}
Using the regularity conditions $\left(  \ref{eq:reg}\right)  $, eqns.
$\left(  \ref{eq:txty}\right)  $ reduce to%
\begin{align}
hA_{t}^{\prime}\left(  r_{h}\right)  h_{ty}\left(  r_{h}\right)  -\left[
\frac{\alpha^{2}}{\mathcal{L}^{\left(  1,0\right)  }\left(  s_{h}%
,p_{h}\right)  }+\frac{h^{2}}{r_{h}^{2}}\right]  h_{tx}\left(  r_{h}\right)
&  =A_{t}^{\prime}\left(  r_{h}\right)  E_{x}+\frac{hE_{y}}{r_{h}^{2}}%
\text{,}\nonumber\\
hA_{t}^{\prime}\left(  r_{h}\right)  h_{tx}\left(  r_{h}\right)  +\left[
\frac{\alpha^{2}}{\mathcal{L}^{\left(  1,0\right)  }\left(  s_{h}%
,p_{h}\right)  }+\frac{h^{2}}{r_{h}^{2}}\right]  h_{ty}\left(  r_{h}\right)
&  =\frac{hE_{x}}{r_{h}^{2}}-A_{t}^{\prime}\left(  r_{h}\right)  E_{y}\text{,}
\label{eq:txtyrh}%
\end{align}
where $s_{h}$ and $p_{h}$ are given by eqns. $\left(  \ref{eq:shph}\right)  $.
Solving eqns. $\left(  \ref{eq:txtyrh}\right)  $ for $h_{ti}\left(
r_{h}\right)  $ in terms of $E_{i}$ and using the regularity conditions
$\left(  \ref{eq:reg}\right)  $ to evaluate eqns. $\left(  \ref{eq:Jxy}%
\right)  $ at $r=r_{h}$, one can relate the currents $\left\langle
\mathcal{J}^{i}\right\rangle $ to the electric fields $E_{i}$ via%
\begin{equation}
\left\langle \mathcal{J}^{x}\right\rangle =\sigma_{xx}E_{x}+\sigma_{xy}%
E_{y}\text{ and }\left\langle \mathcal{J}^{y}\right\rangle =\sigma_{yy}%
E_{y}+\sigma_{yx}E_{x}\text{,}%
\end{equation}
where the DC conductivities $\sigma_{ij}$ are given by%
\begin{align}
\sigma_{xx}  &  =\sigma_{yy}=\alpha^{2}r_{h}^{2}\frac{A_{t}^{\prime2}\left(
r_{h}\right)  r_{h}^{4}+\frac{\alpha^{2}r_{h}^{2}}{\mathcal{L}^{\left(
1,0\right)  }\left(  s_{h},p_{h}\right)  }+h^{2}}{\left(  h^{2}+\frac
{\alpha^{2}r_{h}^{2}}{\mathcal{L}^{\left(  1,0\right)  }\left(  s_{h}%
,p_{h}\right)  }\right)  ^{2}+h^{2}A_{t}^{\prime2}\left(  r_{h}\right)
r_{h}^{4}}\text{,}\nonumber\\
\sigma_{xy}  &  =-\sigma_{yx}=\frac{2r_{h}^{2}\alpha^{2}+\mathcal{L}^{\left(
1,0\right)  }\left(  s_{h},p_{h}\right)  \left[  h^{2}+r_{h}^{4}A_{t}%
^{\prime2}\left(  r_{h}\right)  \right]  }{\left(  h^{2}+\frac{\alpha^{2}%
r_{h}^{2}}{\mathcal{L}^{\left(  1,0\right)  }\left(  s_{h},p_{h}\right)
}\right)  ^{2}+h^{2}A_{t}^{\prime2}\left(  r_{h}\right)  r_{h}^{4}}%
A_{t}^{\prime}\left(  r_{h}\right)  r_{h}^{2}h-\mathcal{L}^{\left(
0,1\right)  }\left(  s_{h},p_{h}\right)  \text{.} \label{eq:DCconductity}%
\end{align}
To express $\sigma_{ij}$ in terms of $\rho$, $h$ and $T$, one needs to solve
eqns. $\left(  \ref{eq:HT}\right)  $ and $\left(  \ref{eq:rho}\right)  $ for
$r_{h}$ and $A_{t}^{\prime}\left(  r_{h}\right)  $ in terms of $\rho$, $h$ and
$T$ and plug the $r_{h}$ and $A_{t}^{\prime}\left(  r_{h}\right)  $
expressions into eqns. $\left(  \ref{eq:DCconductity}\right)  $. Therefore,
$\sigma_{ij}$ are in general functions of the temperature $T$, the charge
density $\rho$, the magnetic field $h$ and the strength of momentum
dissipation $\alpha$. Notice that the conductivities are left invariant under
the separate scaling symmetries given by%
\begin{equation}
T\rightarrow\lambda T\text{, }\alpha\rightarrow\lambda\alpha\text{,
}h\rightarrow\lambda^{2}h\text{, }\rho\rightarrow\lambda^{2}\rho\text{,}%
\end{equation}
for constant $\lambda$. The resistivity matrix is the inverse of the
conductivity matrix:%
\begin{equation}
R_{xx}=R_{yy}=\frac{\sigma_{xx}}{\sigma_{xx}^{2}+\sigma_{yy}^{2}}\text{ and
}R_{xy}=-R_{yx}=-\frac{\sigma_{xy}}{\sigma_{xx}^{2}+\sigma_{yy}^{2}}\text{.}%
\end{equation}

\subsection{Various Limiting Cases}

In section \ref{Sec:Examples}, we will use eqns. $\left(
\ref{eq:DCconductity}\right)  $ to discuss the properties of the DC
conductivities in some NLED models. Before focusing on a specific model, we
now consider some limiting cases of the general formulae for $\sigma_{ij}$ or
$R_{ij}$.

\subsubsection{Weak and Strong Dissipation Limits}

When $\alpha=0$, the system will restore Lorentz invariance. In a Lorentz
invariant theory, it showed \cite{DCC-Hartnoll:2007ai} that the DC
conductivities in the presence of a magnetic field were%
\begin{equation}
\sigma_{xx}=\sigma_{yy}=0\text{ and }\sigma_{xy}=-\sigma_{yx}=\frac{\rho}%
{h}\text{.} \label{eq:LIDC}%
\end{equation}
As a check, we find that, in the weak dissipation limit with $\alpha^{2}\ll1$,
the DC conductivities in eqns. $\left(  \ref{eq:DCconductity}\right)  $ become%
\begin{equation}
\sigma_{xx}=\sigma_{yy}=\frac{\alpha^{2}r_{h}^{2}}{h^{2}}+\mathcal{O}\left(
\alpha^{4}\right)  \text{ and }\sigma_{xy}=-\sigma_{yx}=\frac{\rho}%
{h}+\mathcal{O}\left(  \alpha^{4}\right)  \text{,}%
\end{equation}
which are consistent with eqns. $\left(  \ref{eq:LIDC}\right)  $.

In the strong dissipation limit with $\alpha^{2}\gg1$, we find that the DC
conductivities become%
\begin{equation}
\sigma_{xx}=\sigma_{yy}=\mathcal{L}^{\left(  1,0\right)  }\left(  s_{h}%
,p_{h}\right)  +\mathcal{O}\left(  \alpha^{-2}\right)  \text{ and }\sigma
_{xy}=-\sigma_{yx}=-\mathcal{L}^{\left(  0,1\right)  }\left(  s_{h}%
,p_{h}\right)  +\mathcal{O}\left(  \alpha^{-2}\right)  .
\label{eqn:ProbeLimit}%
\end{equation}
It is noteworthy that eqns. $\left(  \ref{eqn:ProbeLimit}\right)  $ agree with
the results in \cite{IN-Guo:2017bru}, where the DC conductivities were
computed for a probe NLED field. In fact, when $\alpha^{2}\gg1$, the geometry
is almost determined by the contributions from the axionic sector, and hence
the NLED field can be approximated as a probe one.

\subsubsection{Vanishing Magnetic Field and Charge Density}

For the $h=0$ case, the DC conductivities reduce to%
\begin{equation}
\sigma_{xx}=\sigma_{yy}=\frac{\rho^{2}}{r_{h}^{2}\alpha^{2}}+\mathcal{L}%
^{\left(  1,0\right)  }\left(  \frac{A_{t}^{\prime2}\left(  r_{h}\right)  }%
{2},0\right)  \text{ and }\sigma_{xy}=-\sigma_{yx}=-\mathcal{L}^{\left(
0,1\right)  }\left(  \frac{A_{t}^{\prime2}\left(  r_{h}\right)  }{2},0\right)
\text{,}%
\end{equation}
where $A_{t}^{\prime}\left(  r_{h}\right)  $ is obtained by solving
\begin{equation}
\rho=\mathcal{L}^{\left(  1,0\right)  }\left(  \frac{A_{t}^{\prime2}\left(
r_{h}\right)  }{2},0\right)  A_{t}^{\prime}\left(  r_{h}\right)  .
\end{equation}
For the iDBI Lagrangian, our results reduce to eqn. $\left(  3.1\right)  $ in
\cite{IN-Baggioli:2016oju}.

At zero charge density $\rho=0$, the DC conductivities become%
\begin{equation}
\sigma_{xx}^{-1}=\sigma_{yy}^{-1}=\frac{h^{2}}{\alpha^{2}r_{h}^{2}}+\frac
{1}{\mathcal{L}^{\left(  1,0\right)  }\left(  -\frac{h^{2}}{2r_{h}^{4}%
},0\right)  }\text{ and }\sigma_{xy}=-\sigma_{yx}=-\mathcal{L}^{\left(
0,1\right)  }\left(  -\frac{h^{2}}{2r_{h}^{4}},0\right)  \text{.}%
\end{equation}
These DC conductivities are in general non-zero and can be interpreted as
incoherent contributions \cite{DCC-Davison:2015bea}, known as the charge
conjugation symmetric contribution $\sigma_{ccs}$. There is another
contribution from explicit charge density relaxed by some momentum
dissipation, $\sigma_{diss}$, which depends on the charge density $\rho$. Our
results show that, for a general NLED model, the DC conductivities usually
depend on $\sigma_{diss}$ and $\sigma_{ccs}$ in a nontrivial way.

\subsubsection{High Temperature Limit}

\begin{figure}[tb]
\begin{center}
\subfigure[{The red region indicates metallic behavior. The blue region
indicates insulating behavior.}]{
\includegraphics[width=0.45\textwidth]{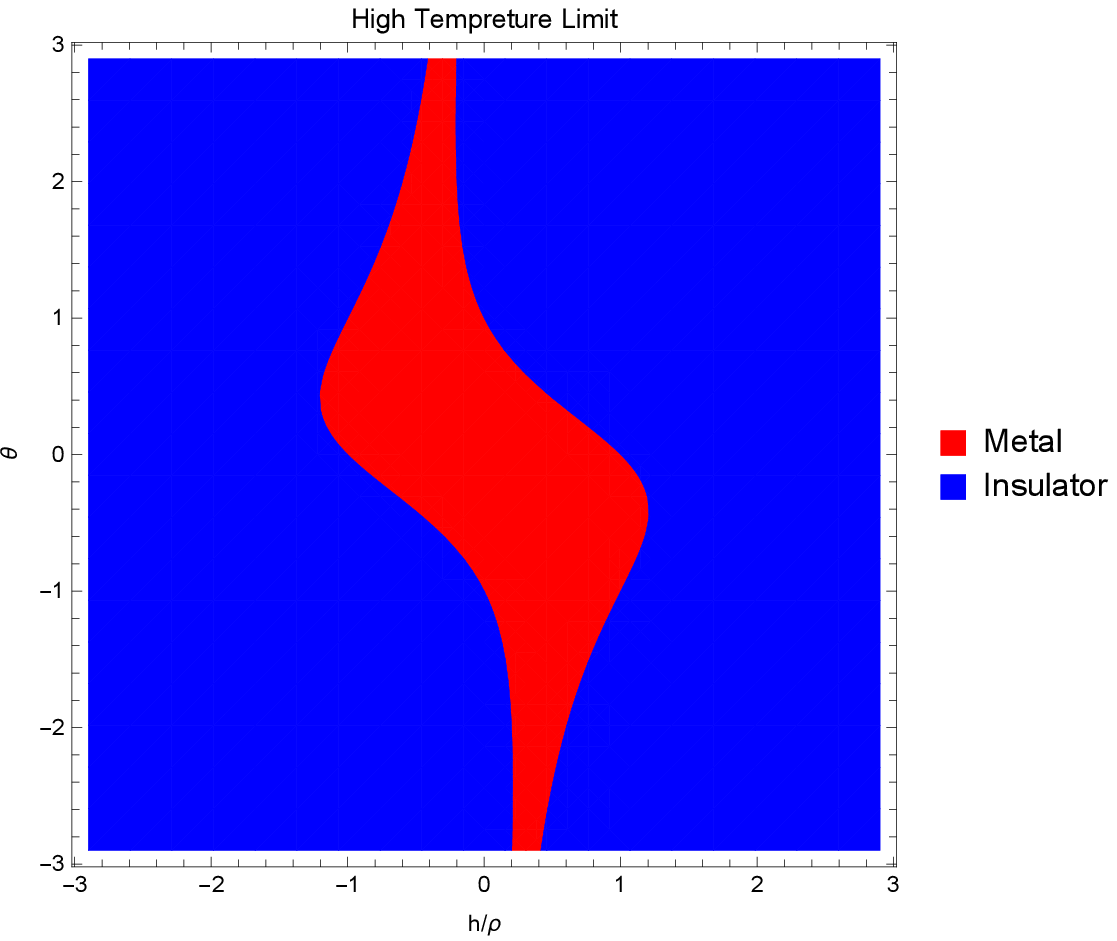}\label{fig:MIT}}
\subfigure[{Yellow region indicates Mott-like behavior. Green Region indicates negative magneto-resistance.}]{
\includegraphics[width=0.45\textwidth]{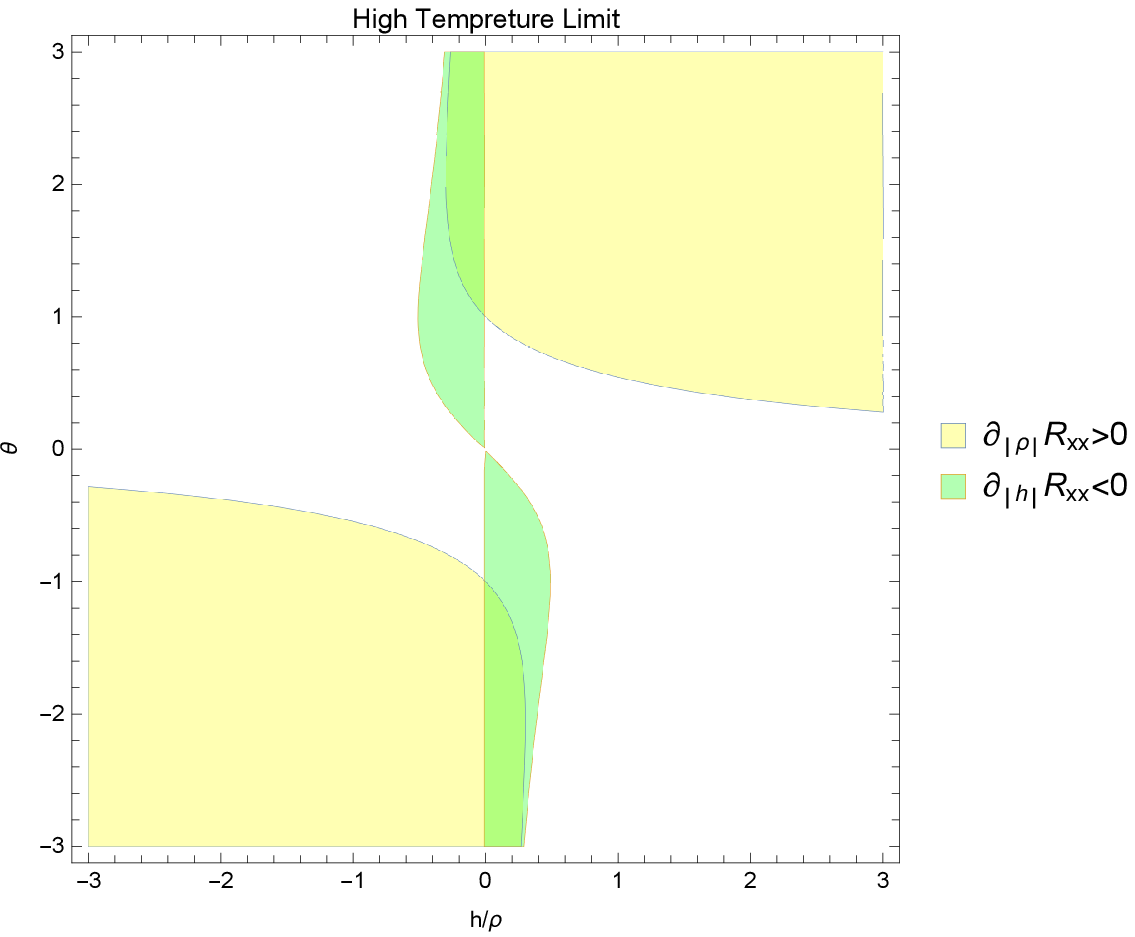}\label{fig:HTMOTTNM}}
\end{center}
\caption{Parameter space in terms of $\theta$ and $h/\rho$ in the high
temperature limit. Note that $\theta\equiv\mathcal{L}^{\left(  0,1\right)
}\left(  0,0\right)  $.}%
\end{figure}

Finally, we consider the high temperature limit $T\gg\left(  \sqrt{h}\text{,
}\sqrt{\rho}\text{, }\alpha\right)  $. In this limit, eqn. $\left(
\ref{eq:HT}\right)  $ gives%
\begin{equation}
T\approx\frac{3}{4\pi}r_{h}.
\end{equation}
The resistance then reduces to%
\begin{equation}
R_{xx}=\frac{1}{1+\theta^{2}}\left\{  1+\frac{9}{16\pi^{2}\alpha^{2}T^{2}%
}\left[  \left(  1+\theta^{2}\right)  h^{2}+2\theta\rho h-\frac{1-\theta^{2}%
}{1+\theta^{2}}\rho^{2}\right]  \right\}  +\mathcal{O}\left(  T^{-4}\right)
\text{,} \label{eq:Rxx}%
\end{equation}
which only depends on $\theta=\mathcal{L}^{\left(  0,1\right)  }\left(
0,0\right)  $ and is independent of the nonlinear effects of the NLED field.
This is understood as the nonlinear terms being suppressed by the high
temperature. One can define a metal and an insulator for $dR_{xx}/dT>0$ and
$dR_{xx}/dT<0$, respectively. Eqn. $\left(  \ref{eq:Rxx}\right)  $ shows that,
for any NLED model in the high temperature, a metal-insulator transition (MIT)
occurs when the term $\left(  1+\theta^{2}\right)  h^{2}+2\theta\rho
h-\frac{1-\theta^{2}}{1+\theta^{2}}\rho^{2}$ changes the sign. In FIG.
\ref{fig:MIT}, we plot the parameter space for a metal and an insulator with
respect to $h/\rho$ and $\theta$. Note that, if $\theta\neq0$, there is no
$\left(  \rho,h\right)  \rightarrow\left(  \rho,-h\right)  $ or $\left(
\rho,h\right)  \rightarrow\left(  -\rho,h\right)  $ symmetries for
$\sigma_{ij}$ or $R_{ij}$. However, $\sigma_{ij}$ or $R_{ij}$ are invariant
under $\left(  \rho,h\right)  \rightarrow\left(  -\rho,-h\right)  $. The
parameter space for $\partial_{\left\vert \rho\right\vert }R_{xx}$ and
$\partial_{\left\vert h\right\vert }R_{xx}$ are plotted in FIG.
\ref{fig:HTMOTTNM}, where we find

\begin{itemize}
\item Green Region: In this region, one has that $\partial R_{xx}%
/\partial\left\vert h\right\vert <0$. To describe how the electrical
resistance responds to an externally-applied magnetic field, one can define
magneto-resistance as
\begin{equation}
MR=\frac{R_{xx}\left(  h\right)  -R_{xx}\left(  0\right)  }{R_{xx}\left(
0\right)  }\text{.}%
\end{equation}
So\ the green region has negative magneto-resistance at given temperature and
charge density.

\item Yellow Region: In this region, one has that $\partial R_{xx}%
/\partial\left\vert \rho\right\vert >0$. This is Mott-like behavior, which can
be explained by the electronic traffic jam: strong enough $e$-$e$ interactions
prevent the available mobile charge carriers to efficiently transport charges.
In particular, when $h=0$, eqn. $\left(  \ref{eq:Rxx}\right)  $ gives that
$\partial R_{xx}/\partial\left\vert \rho\right\vert >0$ as long as $\theta
^{2}>1$.
\end{itemize}

If the NLED Lagrangian $\mathcal{L}\left(  s,p\right)  $ is CP invariant, one
has $\theta=0$. In this case, $R_{xx}$ becomes%
\begin{equation}
R_{xx}=1+\frac{9\left(  h^{2}-\rho^{2}\right)  }{16\pi^{2}\alpha^{2}T^{2}%
}+\mathcal{O}\left(  T^{-4}\right)  \text{,} \label{eq:Rxxtheta=0}%
\end{equation}
which gives that the system displays metallic behavior for $\left\vert
h/\rho\right\vert <1$ and insulating behavior for $\left\vert h/\rho
\right\vert >1$.\textbf{ }Moreover, one always has that $\partial
R_{xx}/\partial\left\vert h\right\vert >0$ and $\partial R_{xx}/\partial
\left\vert \rho\right\vert <0$. Therefore, there is no negative
magneto-resistance or Mott-like behavior for CP invariant NLED models in the
high temperature limit. Note that, in \cite{IN-Cremonini:2017qwq}, eqn.
$\left(  \ref{eq:Rxxtheta=0}\right)  $ was also obtained for the high
temperature limit of the DBI model.

\section{Examples}

\label{Sec:Examples}

In this section, we will use eqns. $\left(  \ref{eq:HT}\right)  $, $\left(
\ref{eq:rho}\right)  $ and $\left(  \ref{eq:DCconductity}\right)  $ to study
the dependence of the in-plane resistance $R_{xx}$ on the temperature $T$, the
charge density $\rho$ and the magnetic field $h$ in Maxwell,
Maxwell-Chern-Simons, Born-Infeld, square and\ logarithmic electrodynamics.
The behavior of $R_{xx}$ in the high temperature limit has already been
discussed in section \ref{Sec:DCC}. So we will focus on the behavior of
$R_{xx}$ around $T=0$ in this section.

\subsection{Maxwell Electrodynamics}

\label{Sec:Maxwell}

\begin{figure}[tb]
\begin{center}
\subfigure[{$T=0$.}]{
\includegraphics[width=0.4\textwidth]{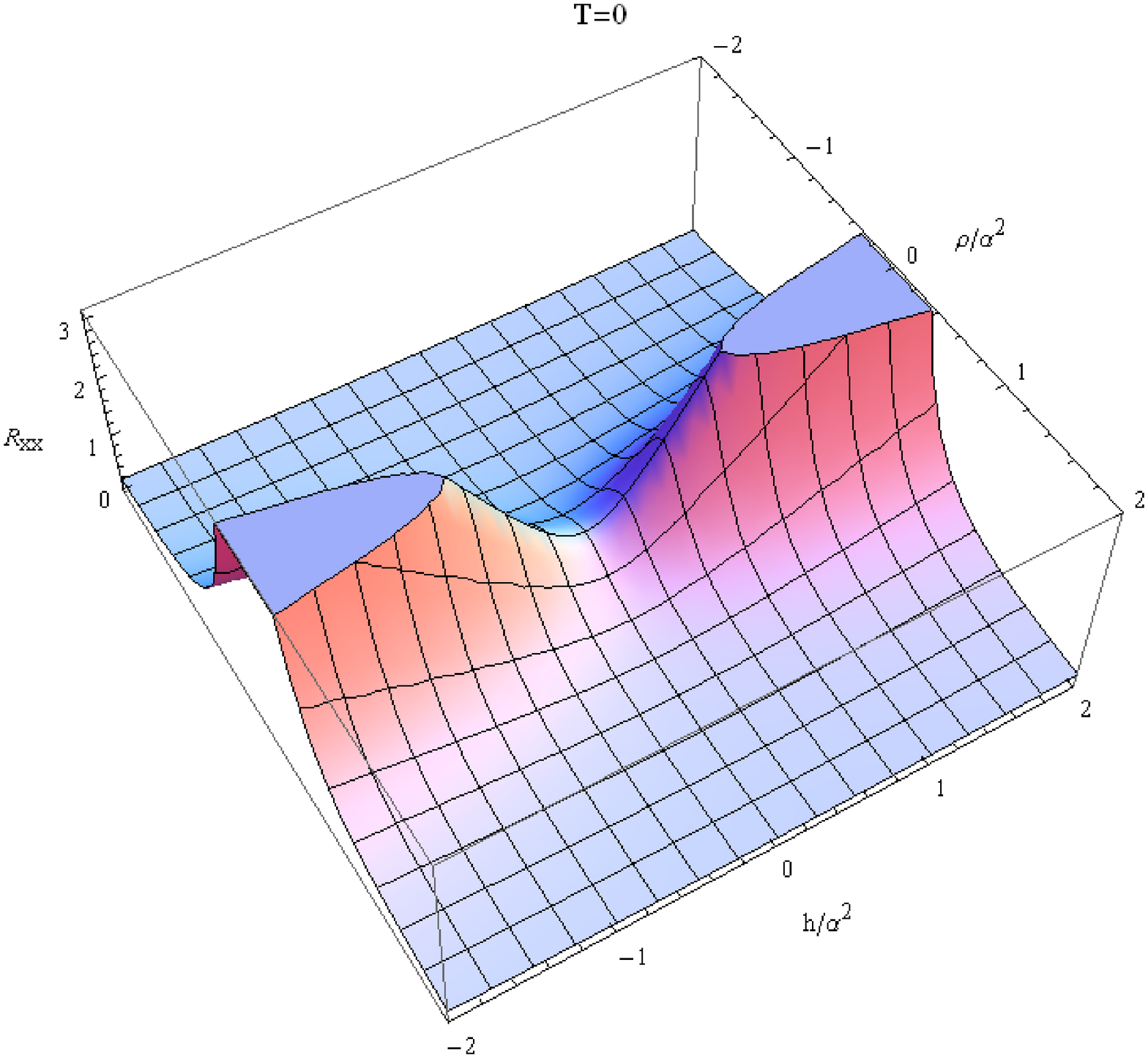}\label{fig:MaxwellT=0}}
\subfigure[{$T/\alpha=1$.}]{
\includegraphics[width=0.4\textwidth]{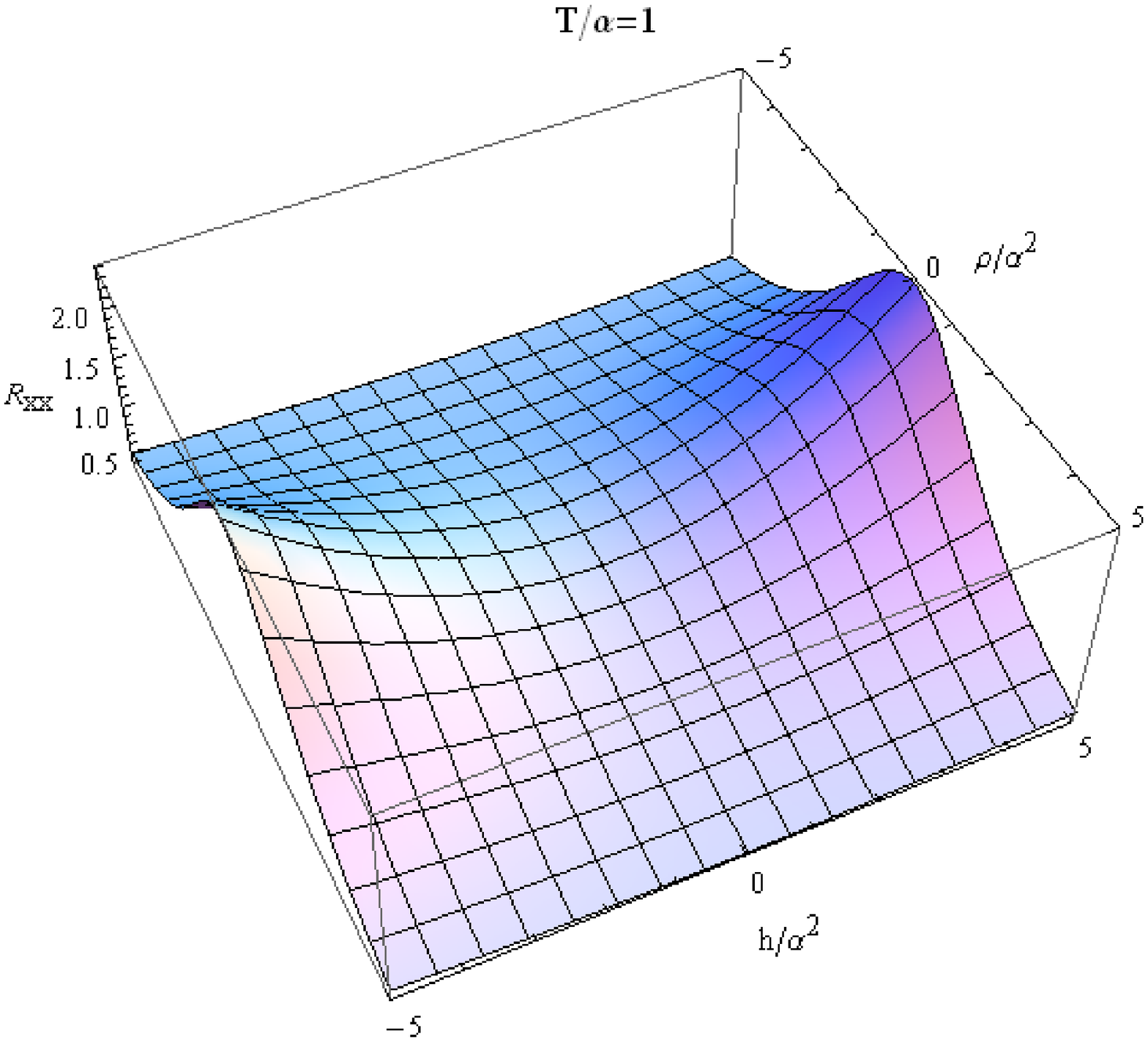}\label{fig:MaxwellTFinite}}
\end{center}
\caption{The resistance $R_{xx}$ versus $\rho/a^{2}$ and $h/a^{2}$ at $T=0$
and $T/\alpha=1$ for Maxwell electrodynamics.}%
\end{figure}

To study the effects of the nonlinear and $\theta$ terms on $R_{xx}$, we first
consider Maxwell electrodynamics,\ in which $\mathcal{L}\left(  s,p\right)
=s$. At $T=0$, the resistance $R_{xx}$ is given by
\begin{equation}
R_{xx}=\frac{\sqrt{1+12\left(  h^{2}/\alpha^{4}+\rho^{2}/\alpha^{4}\right)  }%
}{1+12\rho^{2}/\alpha^{4}},
\end{equation}
which is plotted against $\rho/\alpha^{2}$ and $h/\alpha^{2}$ in FIG.
\ref{fig:MaxwellT=0}. At $T/\alpha=1$, we also plot $R_{xx}$ against
$\rho/\alpha^{2}$ and $h/\alpha^{2}$ in FIG. \ref{fig:MaxwellTFinite}. Both
figures show the saddle surfaces, which imply that $\partial R_{xx}%
/\partial\left\vert h\right\vert >0$ and $\partial R_{xx}/\partial\left\vert
\rho\right\vert <0$. So for Maxwell electrodynamics, $R_{xx}$ does not possess
negative magneto-resistance or Mott-like behavior.

\begin{figure}[tb]
\begin{center}
\subfigure[{Plot of $R_{xx}$ against $h/\rho$ and $T/\sqrt{\rho}$ for $\alpha/\sqrt{\rho}=1$.}]{
\includegraphics[width=0.35\textwidth]{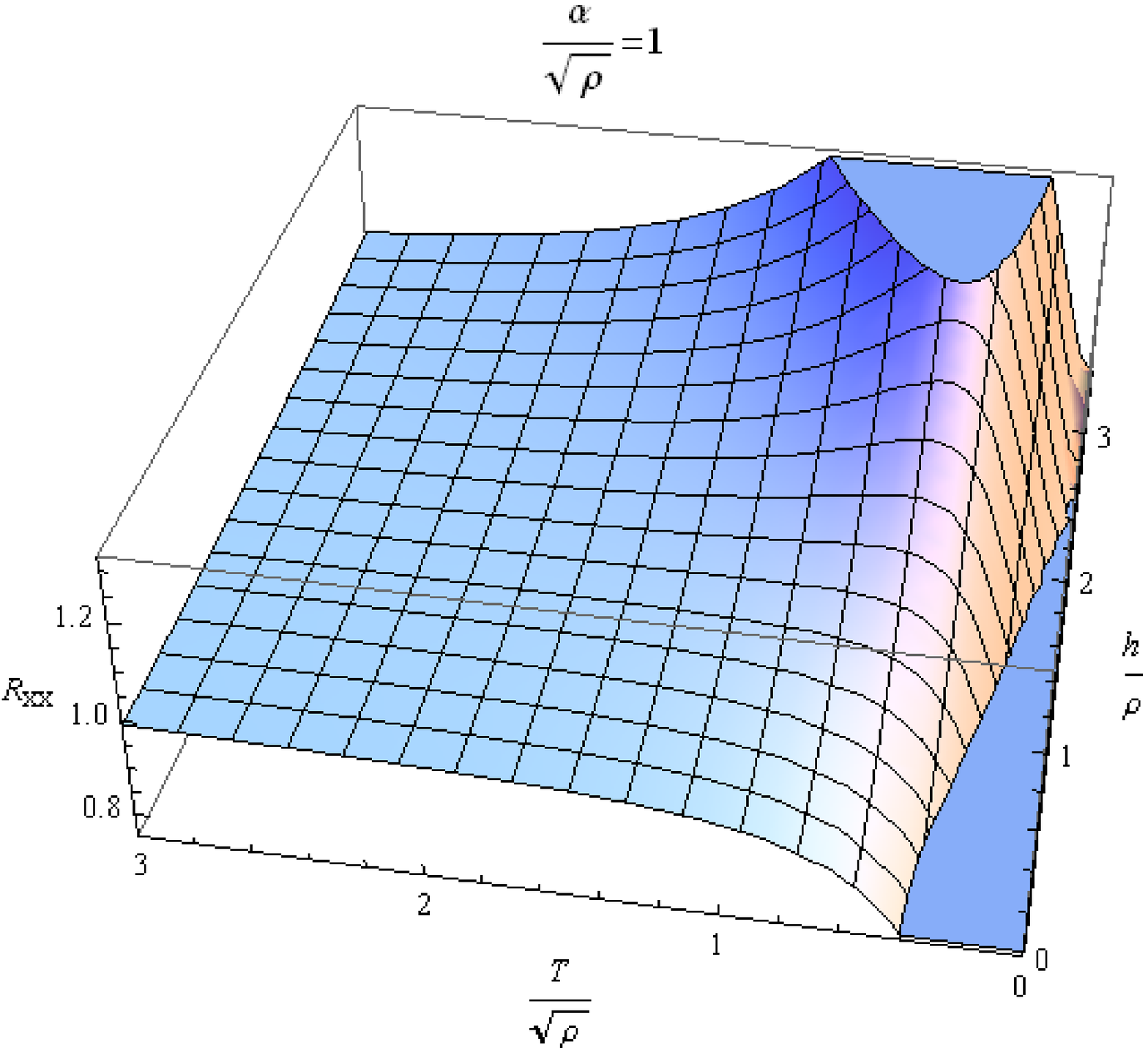}\label{fig:MaxwellMITsmall3D}}
\subfigure[{Plot of
$R_{xx}$ against $T/\sqrt{\rho}$ for various values of $h/\rho$ for $\alpha/\sqrt{\rho}=1$.}]{
\includegraphics[width=0.45\textwidth]{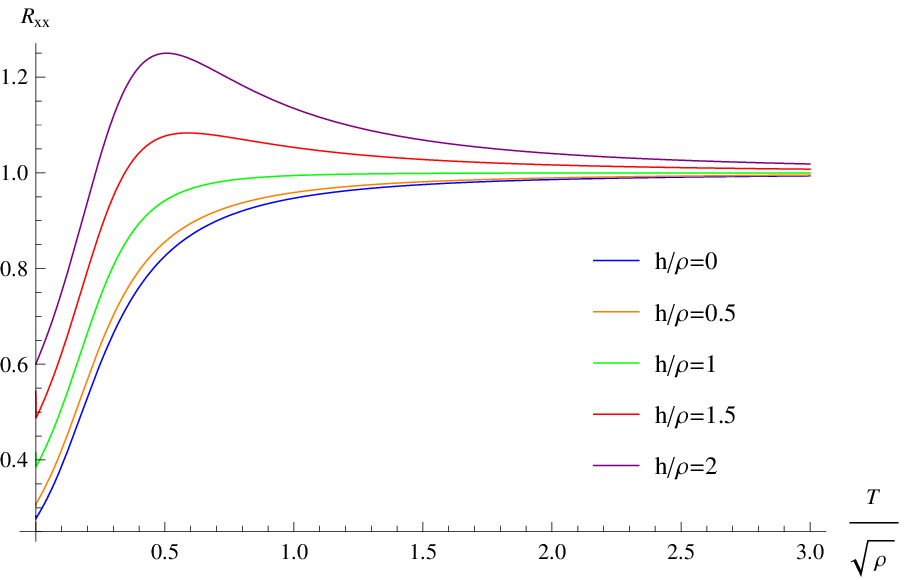}\label{fig:MaxwellMITsmall2D}}
\subfigure[{Plot of $R_{xx}$ against $h/\rho$ and $T/\sqrt{\rho}$ for $\alpha/\sqrt{\rho}=4$.}]{
\includegraphics[width=0.35\textwidth]{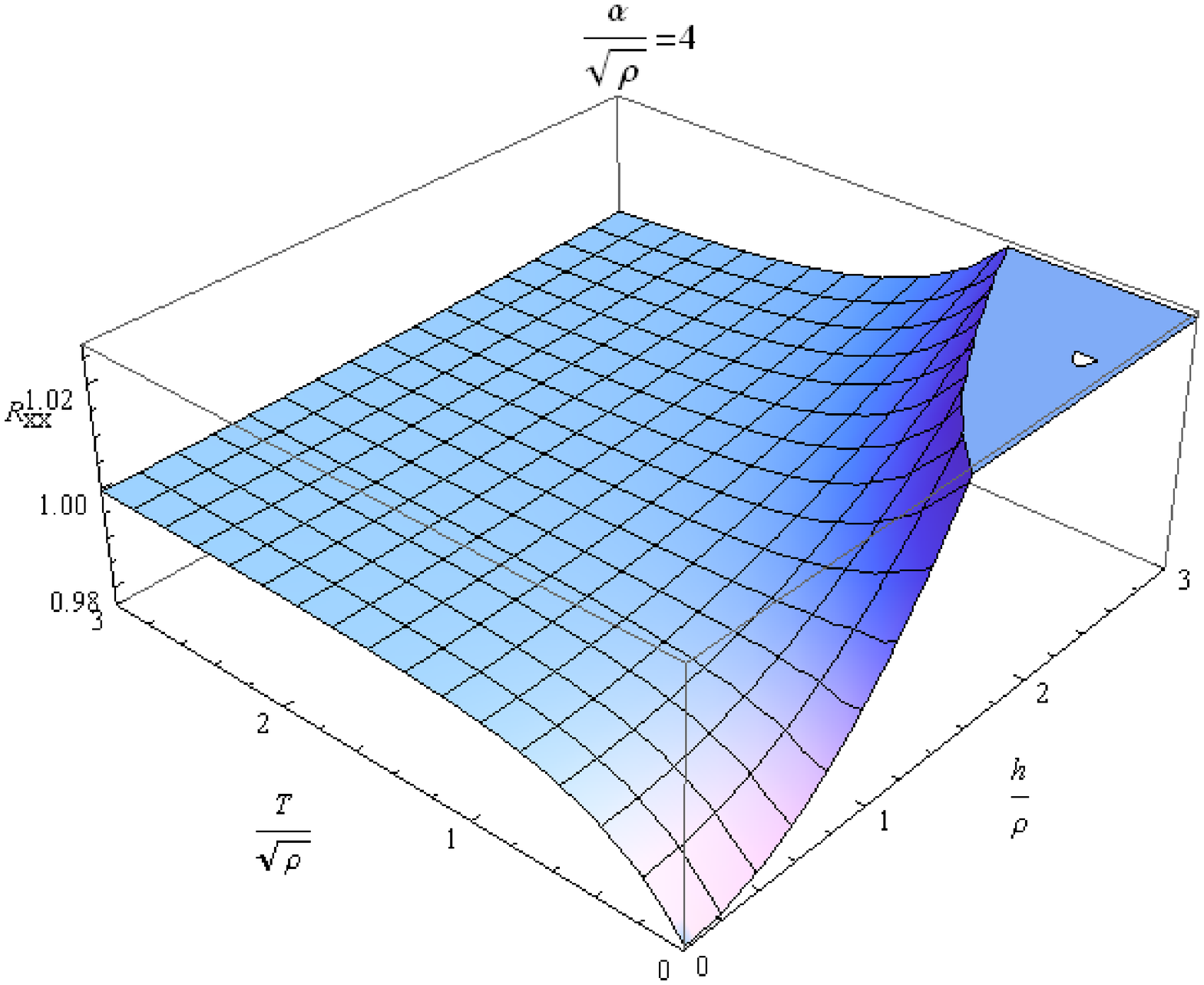}\label{fig:MaxwellMITlarge3D}}
\subfigure[{Plot of
$R_{xx}$ against $T/\sqrt{\rho}$ for various values of $h/\rho$ for $\alpha/\sqrt{\rho}=4$.}]{
\includegraphics[width=0.45\textwidth]{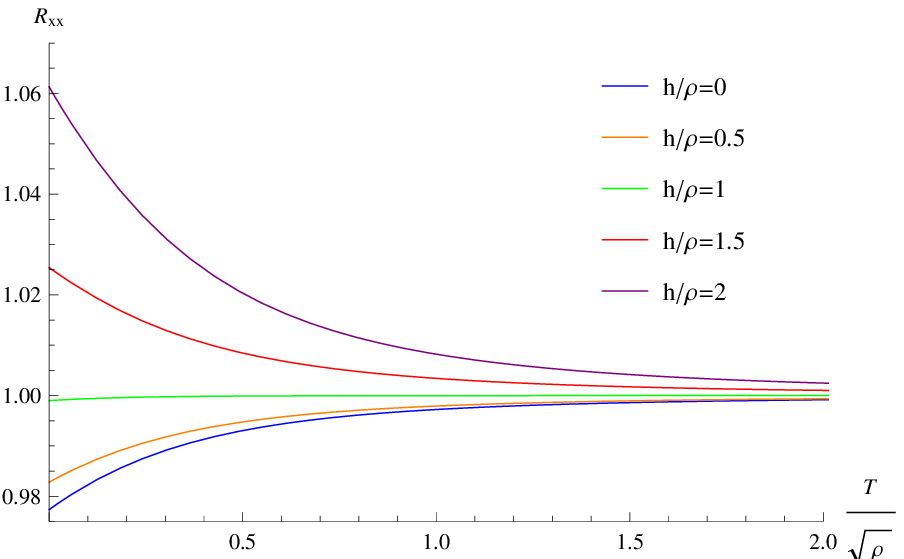}\label{fig:MaxwellMITlarge2D}}
\end{center}
\caption{Plots of $R_{xx}$ with $\alpha/\sqrt{\rho}=1$ and $\alpha/\sqrt{\rho
}=4$ for Maxwell electrodynamics.}%
\label{fig:MaxwellMIT}%
\end{figure}

In FIG. \ref{fig:MaxwellMIT}, we display the dependence of $R_{xx}$ on
$h/\rho$ and $T/\sqrt{\rho}$ for $\alpha/\sqrt{\rho}=1$ and $\alpha/\sqrt
{\rho}=4$, respectively. For $h<\rho$, FIG. \ref{fig:MaxwellMIT} shows that
the temperature dependence of $R_{xx}$ is similar in both $\alpha/\sqrt{\rho
}=1$ and $\alpha/\sqrt{\rho}=4$ cases. The resistance $R_{xx}$ increases
monotonically as the temperature increasing, which corresponds to metallic
behavior. For $h>\rho$, the $\alpha/\sqrt{\rho}=1$ and $\alpha/\sqrt{\rho}=4$
cases show different temperature dependence of $R_{xx}$. When $\alpha
/\sqrt{\rho}=1$, FIGs. \ref{fig:MaxwellMITsmall3D} and
\ref{fig:MaxwellMITsmall2D} show that, as the temperature increases, $R_{xx}$
increases first and then decreases monotonically after reaching a maximum. The
insulating behavior appears at high temperatures in this case. When
$\alpha/\sqrt{\rho}=4$, FIGs. \ref{fig:MaxwellMITlarge3D} and
\ref{fig:MaxwellMITlarge2D} show that $R_{xx}$ decreases monotonically as one
increases the temperature, which corresponds to insulating behavior. So in the
$\alpha/\sqrt{\rho}=4$ and $\alpha/\sqrt{\rho}=1$ cases at high temperatures,
increasing the magnetic field would induce a finite-temperature transition or
crossover from metallic to insulating behavior.

\subsection{Maxwell-Chern-Simons Electrodynamics}

\begin{figure}[tb]
\begin{center}
\subfigure[{Plot of $R_{xx}$ against $h/\alpha^{2}$ and $\rho/\alpha^{2}$ for $\theta=3$.}]{
\includegraphics[width=0.35\textwidth]{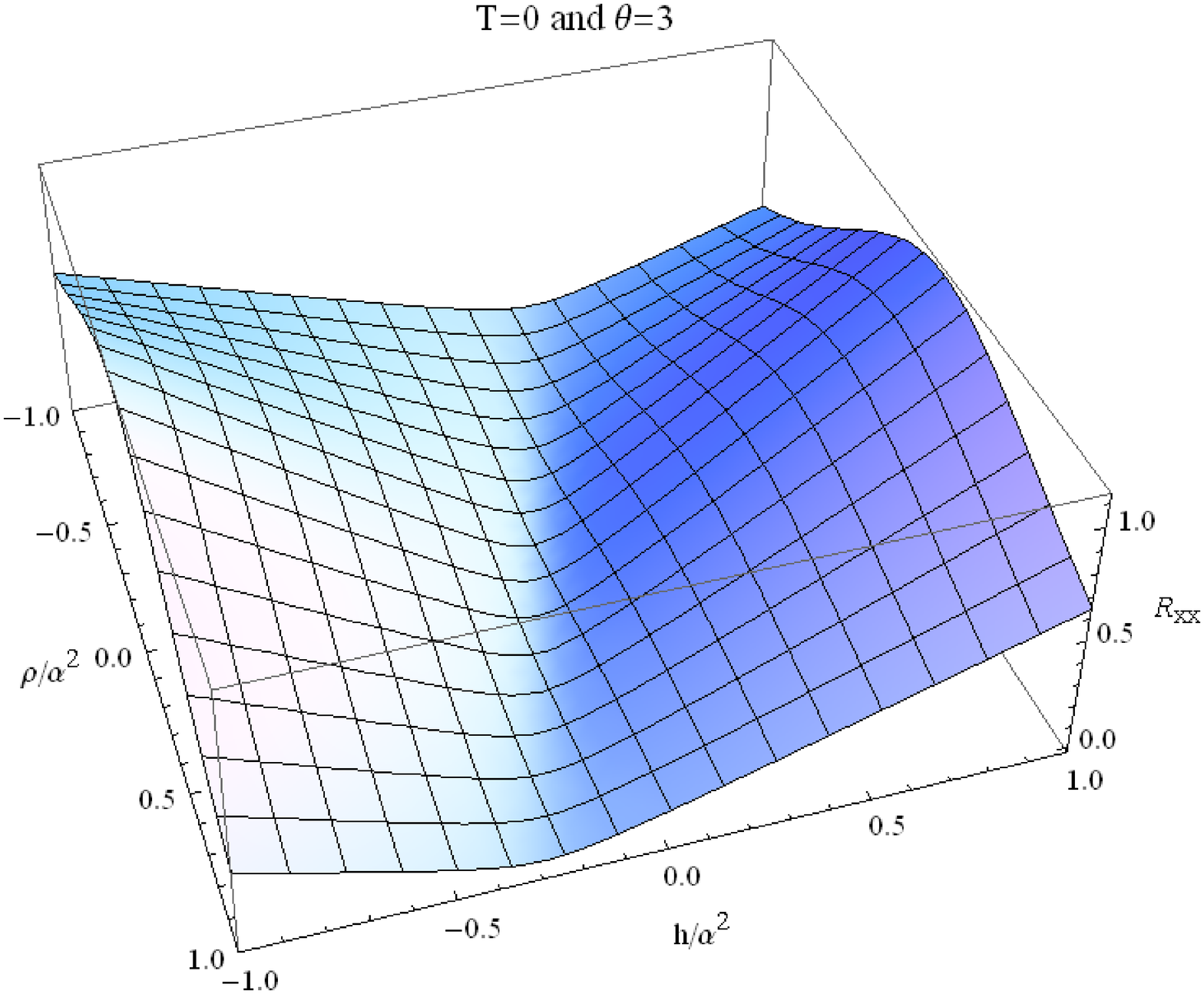}\label{fig:MCST0th3}}
\subfigure[{Plot of $R_{xx}$ against $\rho/\alpha^{2}$ for various values of $h
/\alpha^{2}$ for $\theta=3$.}]{
\includegraphics[width=0.4\textwidth]{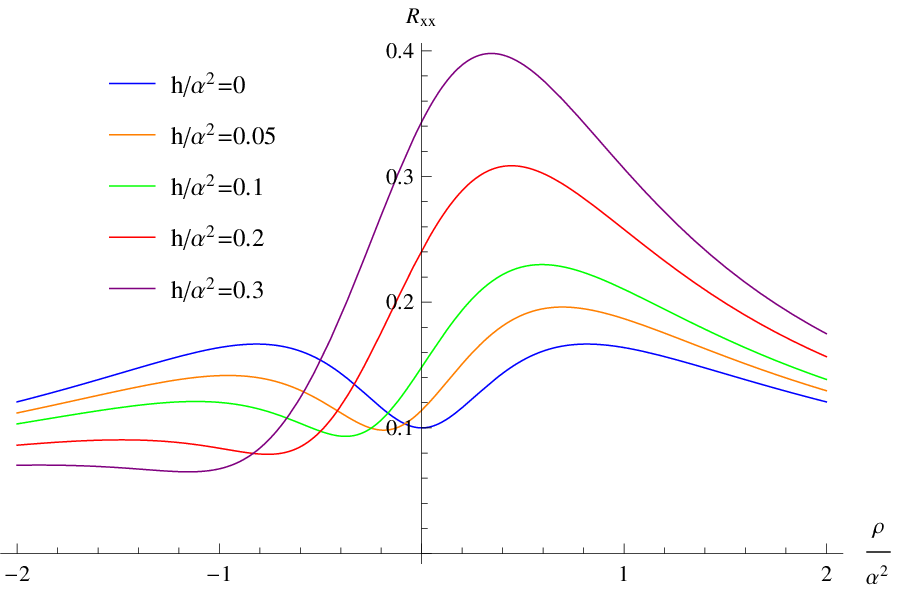}\label{fig:MCST0th32D}}
\end{center}
\caption{Plots of $R_{xx}$ with $\theta=3$ at $T=0$ for Maxwell-Chern-Simons
electrodynamics.}%
\end{figure}

The Lorentz and gauge invariance allow the electrodynamics Lagrangian to have
a CP-violating $\theta$ term%
\begin{equation}
\mathcal{L}\left(  s,p\right)  =s+\theta p\text{.}%
\end{equation}
We now discuss the dependence of $R_{xx}$ on $\rho$ and $h$ at $T=0$. The
resistance $R_{xx}$ can be expressed in terms of $r_{h}$, $\rho$ and $h$:
\begin{equation}
R_{xx}=\frac{\alpha^{2}r_{h}^{2}\left[  \left(  1+\theta^{2}\right)
h^{2}+\alpha^{2}r_{h}^{2}+2\theta h\rho+\rho^{2}\right]  }{\left(
1+\theta^{2}\right)  \alpha^{4}r_{h}^{4}+h^{2}\rho^{2}+2\alpha^{2}r_{h}%
^{2}\rho^{2}+\rho^{2}\left(  \rho+\theta h\right)  ^{2}}.
\end{equation}
At zero temperature, the resistance $R_{xx}$ becomes%
\begin{equation}
R_{xx}=\frac{\sqrt{1+12\left[  h^{2}/\alpha^{4}+\left(  \rho/\alpha^{2}+\theta
h^{2}/\alpha^{2}\right)  ^{2}\right]  }}{1+\theta^{2}+12\rho^{2}/\alpha^{4}}.
\end{equation}
In FIG. \ref{fig:MCST0th3}, we plot $R_{xx}$ versus $\rho/\alpha^{2}$ and
$h^{2}/\alpha^{2}$ with $\theta=3$. Similar to Maxwell electrodynamics, we
have a saddle surface. However, the valley in FIG. \ref{fig:MCST0th3} is at
$\rho/\alpha^{2}+3h^{2}/\alpha^{2}=0$, instead of $\rho/\alpha^{2}=0$. This
twist of the valley would result in the appearance of negative
magneto-resistance and Mott-like behavior.

\begin{figure}[tb]
\begin{center}
\subfigure[{$\theta=1$.}]{
\includegraphics[width=0.4\textwidth]{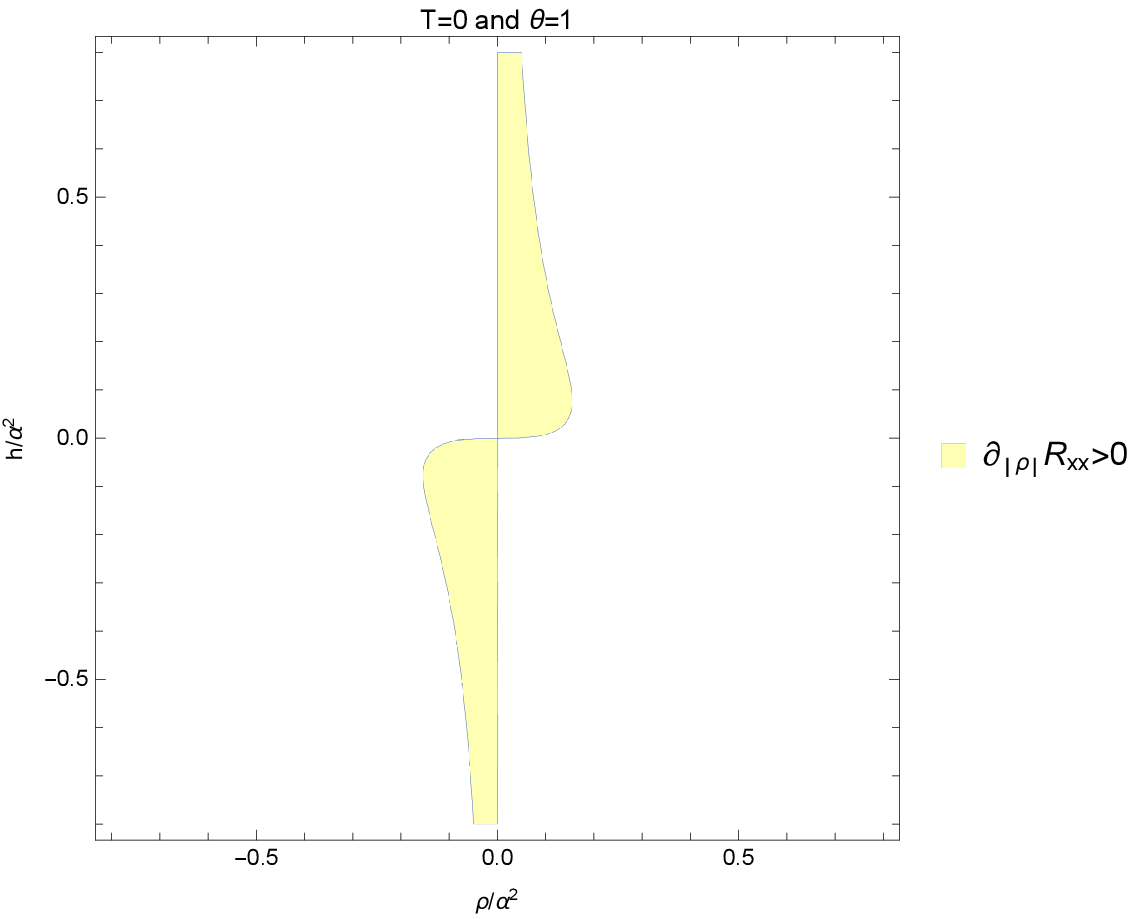}\label{fig:MOTTCS1}}
\subfigure[{$\theta=3$.}]{
\includegraphics[width=0.4\textwidth]{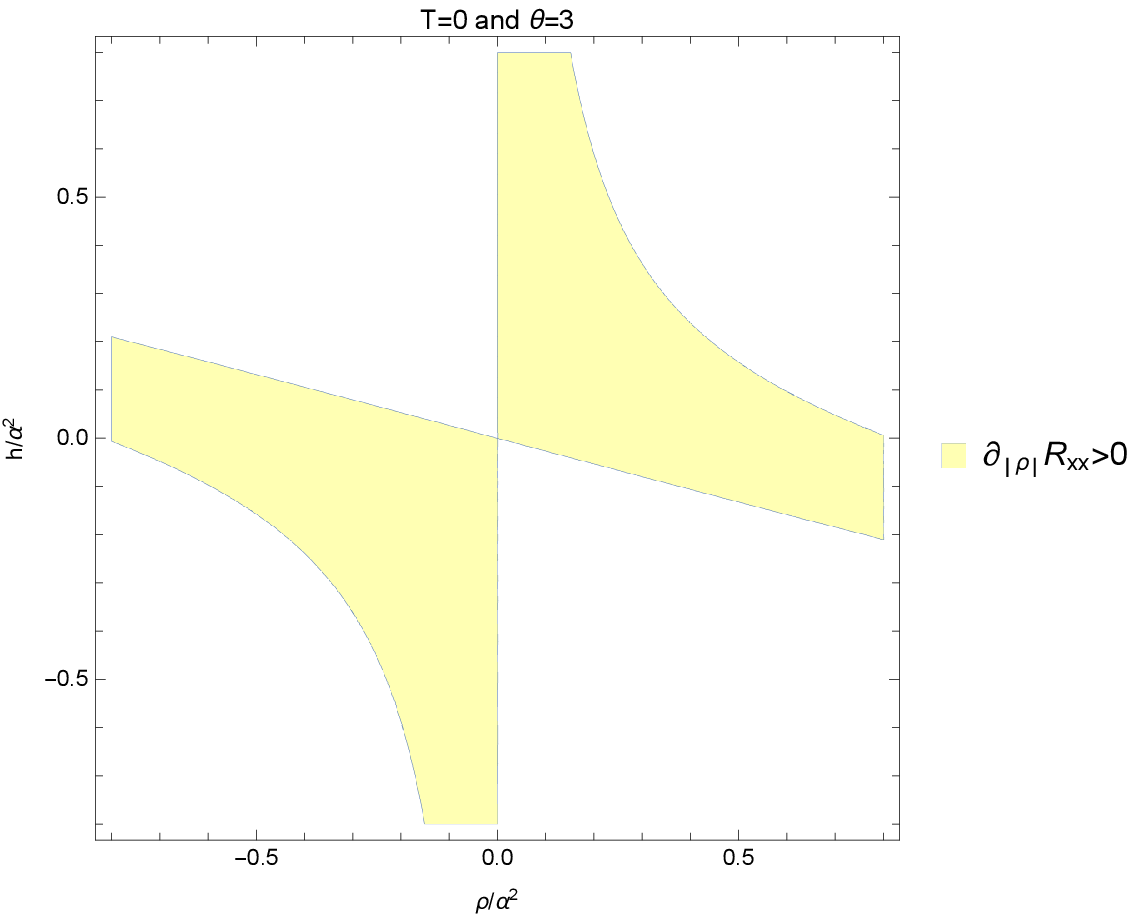}\label{fig:MOTTCS3}}
\end{center}
\caption{Parameter space for $\partial_{\left\vert \rho\right\vert }R_{xx}>0$
in terms of $\rho/\alpha^{2}$ and $h/\alpha^{2}$ with $\theta=1$ and
$\theta=3$ at $T=0$ for Maxwell-Chern-Simons electrodynamics.}%
\label{fig:MOTTCS}%
\end{figure}

The dependence of $R_{xx}$ on $h$ can be obtained by computing $\partial
_{h}R_{xx}$. We find that solving $\partial_{\left\vert h\right\vert }%
R_{xx}<0$ gives%
\begin{equation}
-\frac{\theta}{1+\theta^{2}}<h/\rho<0\text{ for }\theta>0\text{, and }%
0<h/\rho<-\frac{\theta}{1+\theta^{2}}\text{ for }\theta<0\text{,}
\label{eq:csnm}%
\end{equation}
where one has negative magneto-resistance. Note that, in the high temperature
limit, $\partial_{\left\vert h\right\vert }R_{xx}<0$ reduces to%
\begin{equation}
-\frac{2\theta}{1+\theta^{2}}<h/\rho<0\text{ for }\theta>0\text{, and
}0<h/\rho<-\frac{2\theta}{1+\theta^{2}}\text{ for }\theta<0\text{.}%
\end{equation}
When $h=0$, we find that
\begin{equation}
\partial_{\left\vert \rho\right\vert }R_{xx}>0\Rightarrow\rho^{2}<\left(
\frac{\theta^{2}-1}{12}\right)  \alpha^{4}\text{,}%
\end{equation}
which means that there is no Mott-like behavior for $h=0$ if $\theta^{2}\leq
1$. In FIG. \ref{fig:MOTTCS}, we plot the parameter space for $\partial
_{\left\vert \rho\right\vert }R_{xx}>0$ in terms of $\rho/\alpha^{2}$ and
$h/\alpha^{2}$ with $\theta=1$ and $\theta=3$. In the yellow region, one has
$\partial_{\left\vert \rho\right\vert }R_{xx}>0$. As expected, the line $h=0$
is in the yellow region for $\theta=3$ while it is not for $\theta=1$. In FIG.
\ref{fig:MCST0th32D}, we plot $R_{xx}$ versus $\rho/\alpha^{2}$ for various
values of $h/\alpha^{2}$. One has $h/\alpha^{2}=0$ for the blue line, and it
has a minimum at $\rho/\alpha^{2}=0$. As one increases the magnitude of the
charge density, the value of $R_{xx}$ first increases, then reaches a maximum,
and then decreases monotonically. For $h/\alpha^{2}\neq0$, the behavior of
$R_{xx}$ is different when one moves along the positive and negative
$\rho/\alpha^{2}$ directions. Along the positive $\rho/\alpha^{2}$ direction,
the behavior of $R_{xx}$ is similar to the $h/\alpha^{2}=0$ case. However, if
one increases the magnitude of the charge density along the negative
$\rho/\alpha^{2}$ direction, the value of $R_{xx}$ first decreases until
reaching a minimum, then increases until reaching a maximum, and then
decreases monotonically.

\subsection{Born-Infeld Electrodynamics}

\begin{figure}[tb]
\begin{center}
\subfigure[{ $a=1.$}]{
\includegraphics[width=0.32\textwidth]{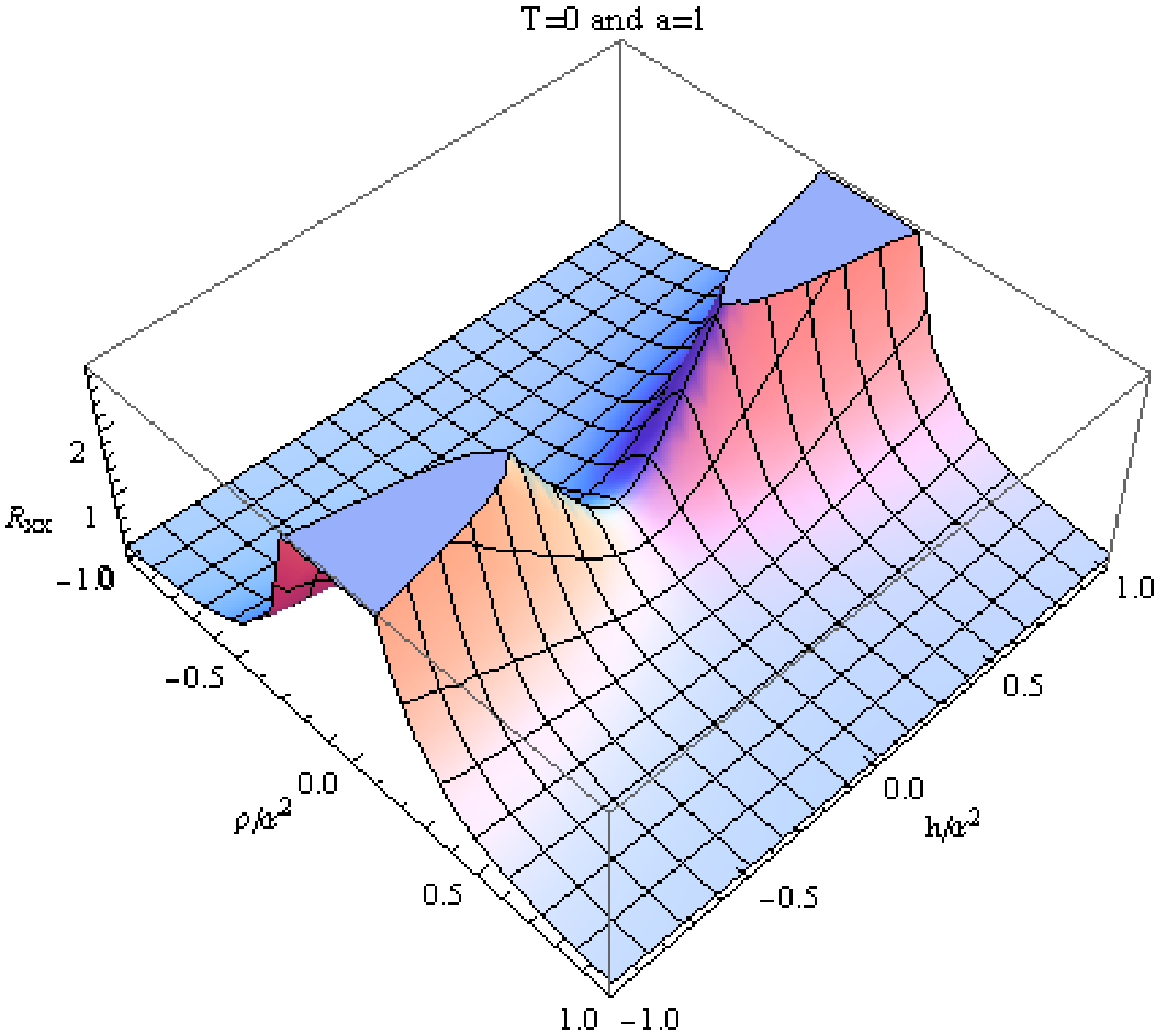}\label{fig:BIP3D}}
\subfigure[{ $a=-0.4.$}]{
\includegraphics[width=0.32\textwidth]{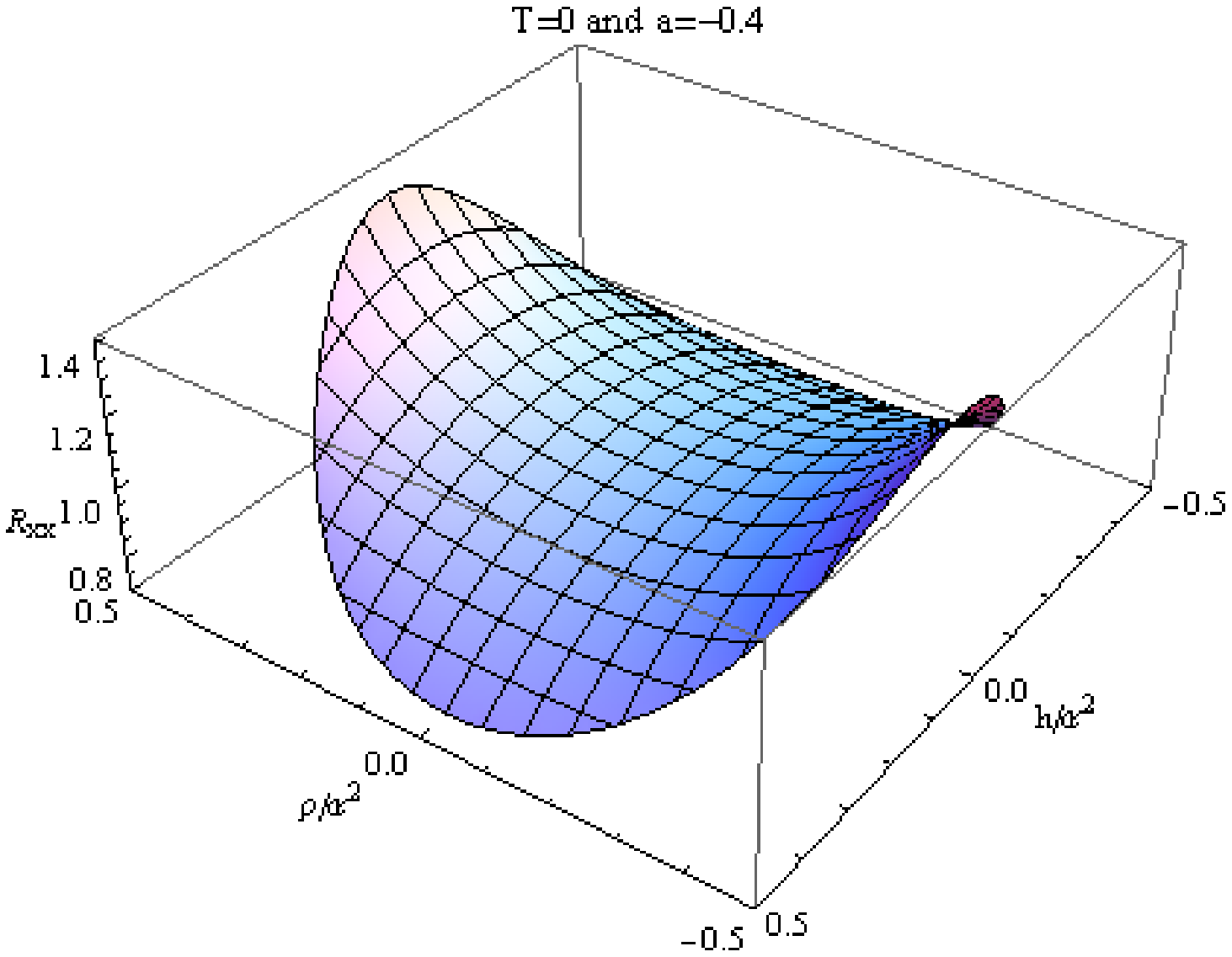}\label{fig:BIMs3D}}
\subfigure[{ $a=-1.$}]{
\includegraphics[width=0.32\textwidth]{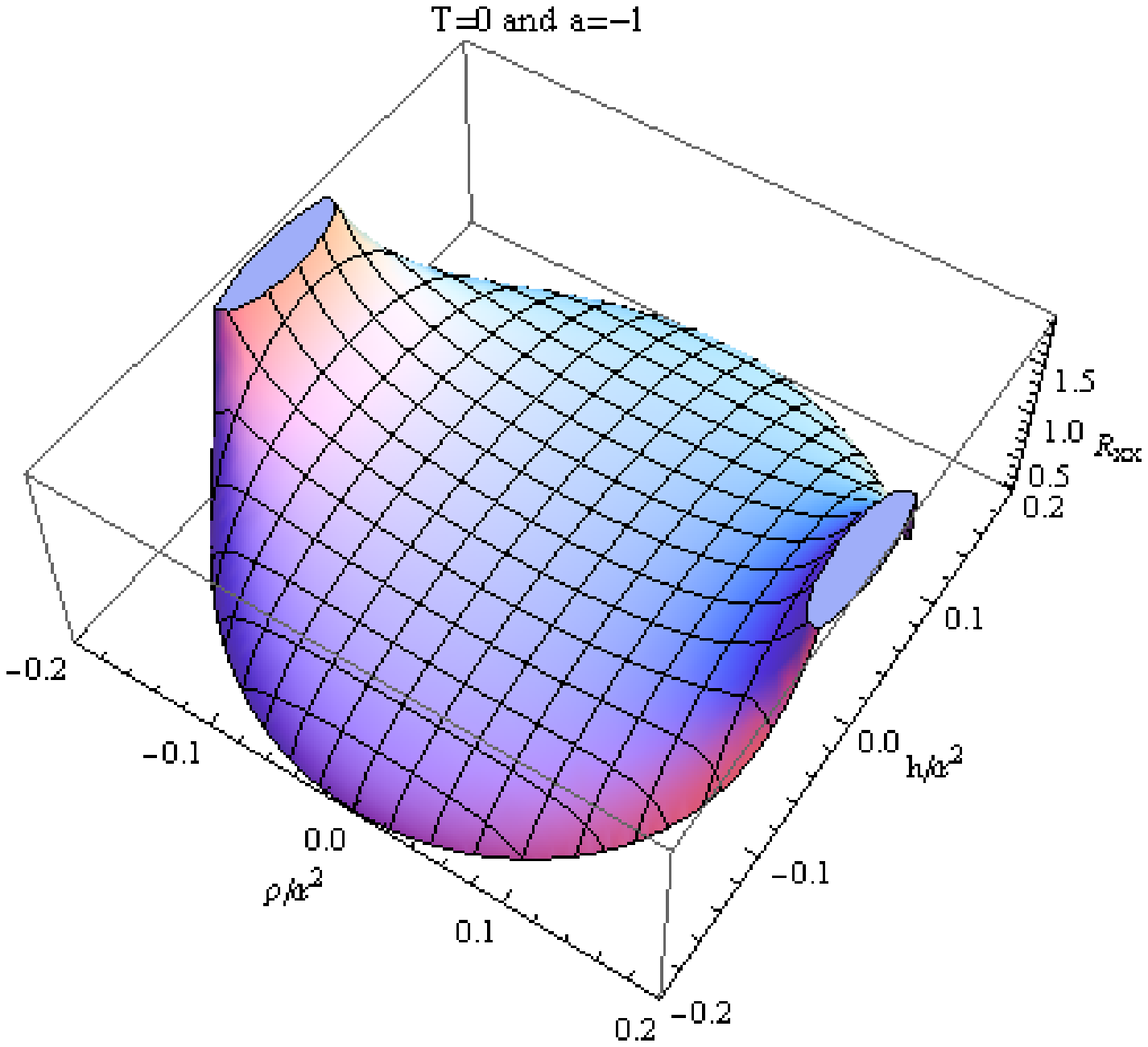}\label{fig:BIMl3D}}
\end{center}
\caption{Plots of $R_{xx}$ versus $\rho/\alpha^{2}$ and $h/\alpha^{2}$ with
$a=1$, $a=-0.4$ and $a=-1$ at $T=0$ for Born-Infeld electrodynamics.}%
\end{figure}

Born-Infeld electrodynamics is described by the Lagrangian density%
\begin{equation}
\mathcal{L}\left(  s,p\right)  =\frac{1}{a}\left(  1-\sqrt{1-2as-a^{2}p^{2}%
}\right)  \text{,} \label{eq:BI}%
\end{equation}
where the coupling parameter $a$ is related to the string tension
$\alpha^{\prime}$ as $a=\left(  2\pi\alpha^{\prime}\right)  ^{2}$. When
$\left\vert a\right\vert \ll1$, we can recover the Maxwell Lagrangian. For the
$a>0$ case, the properties of the resistance $R_{xx}$ were discussed in
\cite{IN-Cremonini:2017qwq}. It showed that the behavior of $R_{xx}$\ obtained
in \cite{IN-Cremonini:2017qwq} was quite similar to that in the Maxwell case,
which has been investigated in section \ref{Sec:Maxwell}. In fact, there is no
appearance of negative magneto-resistance or Mott-like behavior at low
temperatures, maybe for all the temperatures, in both cases. To illustrate
this point, we plot $R_{xx}$ versus $\rho/\alpha^{2}$ and $h^{2}/\alpha^{2}$
at $T=0$ for Born-Infeld electrodynamics with $a=1$ in FIG. \ref{fig:BIP3D},
which shows that $\partial_{\left\vert \rho\right\vert }R_{xx}<0$ and
$\partial_{\left\vert h\right\vert }R_{xx}>0$. Note that FIGs.
\ref{fig:MaxwellT=0} and \ref{fig:BIP3D} look alike. Moreover, both cases have
quite similar behavior of $R_{xx}$ as a function of $h/\rho$ and $T/\sqrt
{\rho}$ for the small and large values of the momentum dissipation parameter.

On the other hand, the $a<0$ case turns out more interesting. For the case
with vanishing magnetic filed, the properties of $\sigma_{xx}$ have been
analyzed in depth in \cite{IN-Baggioli:2016oju}, in which it showed that the
conductivity could decrease with increasing charge density for large enough
self-interaction strength$^{\left[  \ref{ft:1}\right]  }$\footnotetext[1]%
{\label{ft:1} In fact, the NLED Lagrangian $\mathcal{L}\left(  s,p\right)
=\frac{1}{a}\left(  1-\sqrt{1-2as}\right)  $, instead of eqn. $\left(
\ref{eq:BI}\right)  $, was used in \cite{IN-Baggioli:2016oju}. However, for
the $h=0$ case, these two Lagrangian would give the same result for
$\sigma_{xx}$ since they have the same value of $\mathcal{L}^{\left(
1,0\right)  }\left(  s,0\right)  $.}. Here, we extend the analysis to the
non-vanishing magnetic field case. We can solve eqn. $\left(  \ref{eq:Grt}%
\right)  $ for $A_{t}^{\prime}\left(  r\right)  $:%
\begin{equation}
A_{t}^{\prime}\left(  r\right)  =\frac{\rho}{\sqrt{r^{4}+a\left(  h^{2}%
+\rho^{2}\right)  }},
\end{equation}
which shows that there is a singularity at $r=r_{s}\equiv\sqrt[1/4]{-a\left(
h^{2}+\rho^{2}\right)  }$ for $a<0$. To have a physical solution, we need to
hide the singularity behind the horizon: $r_{s}<r_{h}$, which could put an
upper bound on $h^{2}+\rho^{2}$.

\begin{figure}[tb]
\begin{center}
\subfigure[{
Plot of $R_{xx}$ against $\rho/\alpha^{2}$ with $a=-0.4$ at $T=0$ for various
values of $h/\alpha^{2}$.}]{
\includegraphics[width=0.45\textwidth]{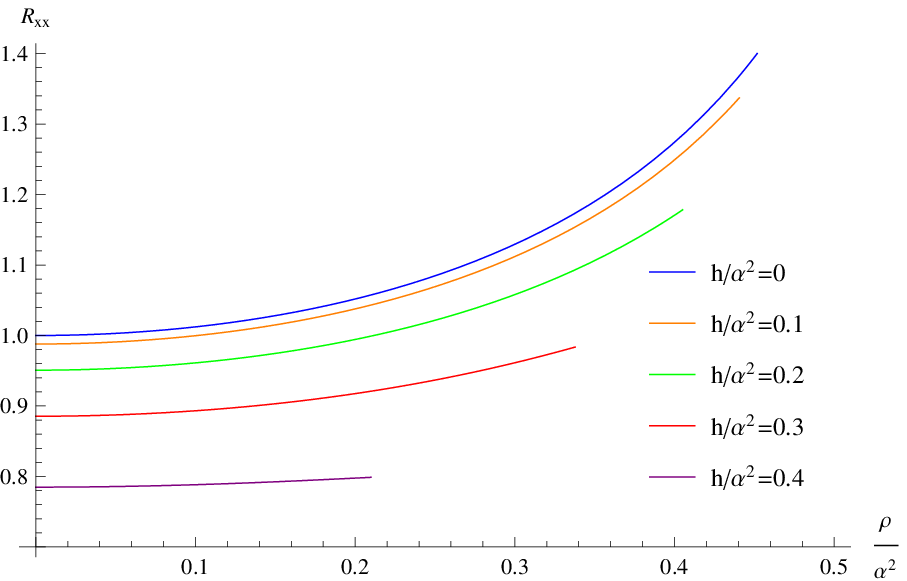}\label{fig:BIMsrho2D}}
\subfigure[{Plot of $R_{xx}$ against $h/\alpha^{2}$ with $a=-0.4$ at $T=0$ for various
values of $\rho/\alpha^{2}$.}]{
\includegraphics[width=0.45\textwidth]{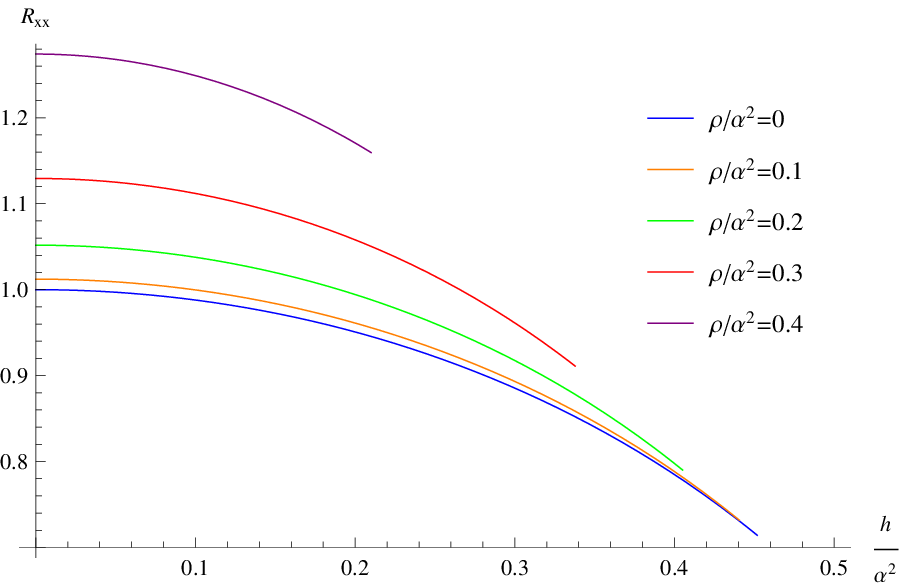}\label{fig:BIMsh2D}}
\subfigure[{Plot of $R_{xx}$ against $\rho/\alpha^{2}$ with $a=-1$ at $T=0$ for various
values of $h/\alpha^{2}$.}]{
\includegraphics[width=0.45\textwidth]{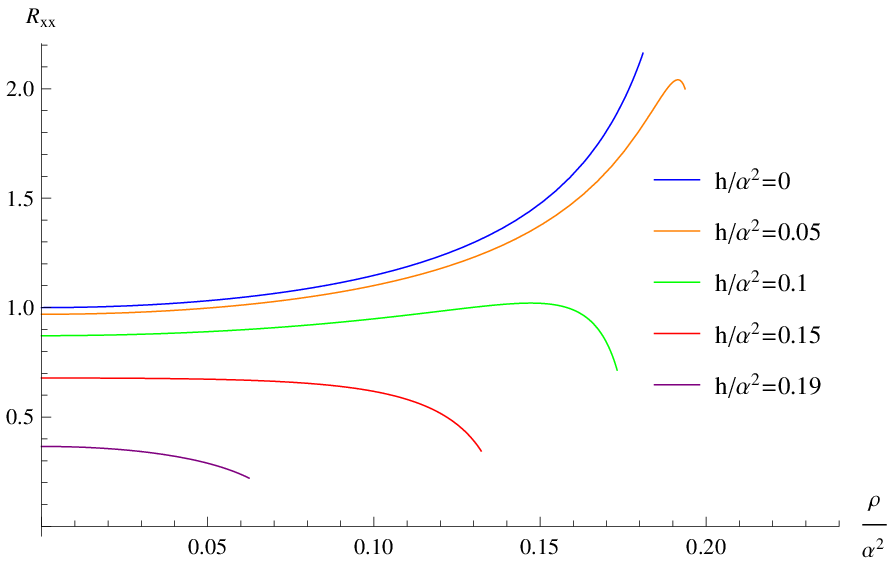}\label{fig:BIMlrho2D}}
\subfigure[{Plot of $R_{xx}$ against $h/\alpha^{2}$ with $a=-1$ at $T=0$ for various
values of $\rho/\alpha^{2}$.}]{
\includegraphics[width=0.45\textwidth]{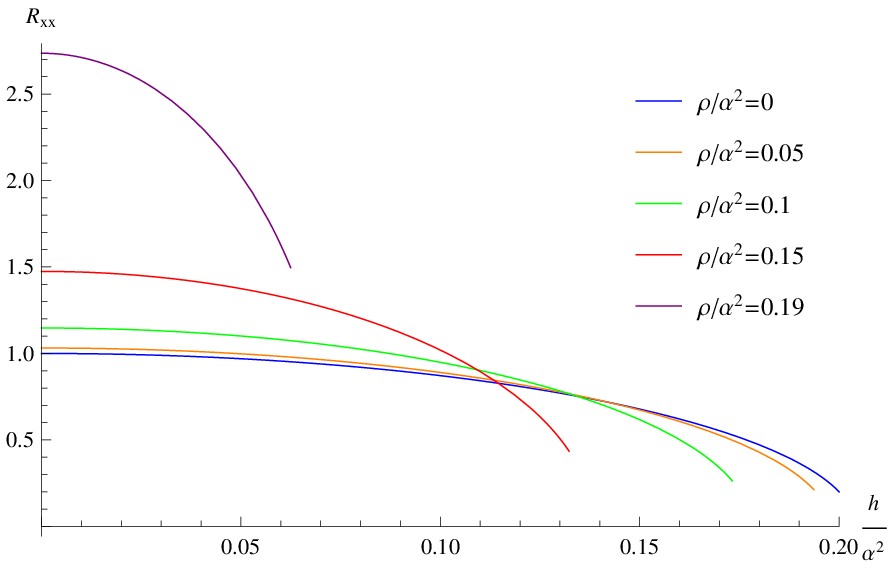}\label{fig:BIMlh2D}}
\end{center}
\caption{Plots of $R_{xx}$ against the charge density $\rho$ and the magnetic
field $h$ with $a=-0.4$ and $a=-1$ at $T=0$ for Born-Infeld electrodynamics.}%
\label{fig:BIMhrho2D}%
\end{figure}

\begin{figure}[tb]
\begin{center}
\subfigure[{Plot of $R_{xx}$ against $h/\rho$ and $T/\sqrt{\rho}$ for $\alpha/\sqrt{\rho}=1$.}]{
\includegraphics[width=0.35\textwidth]{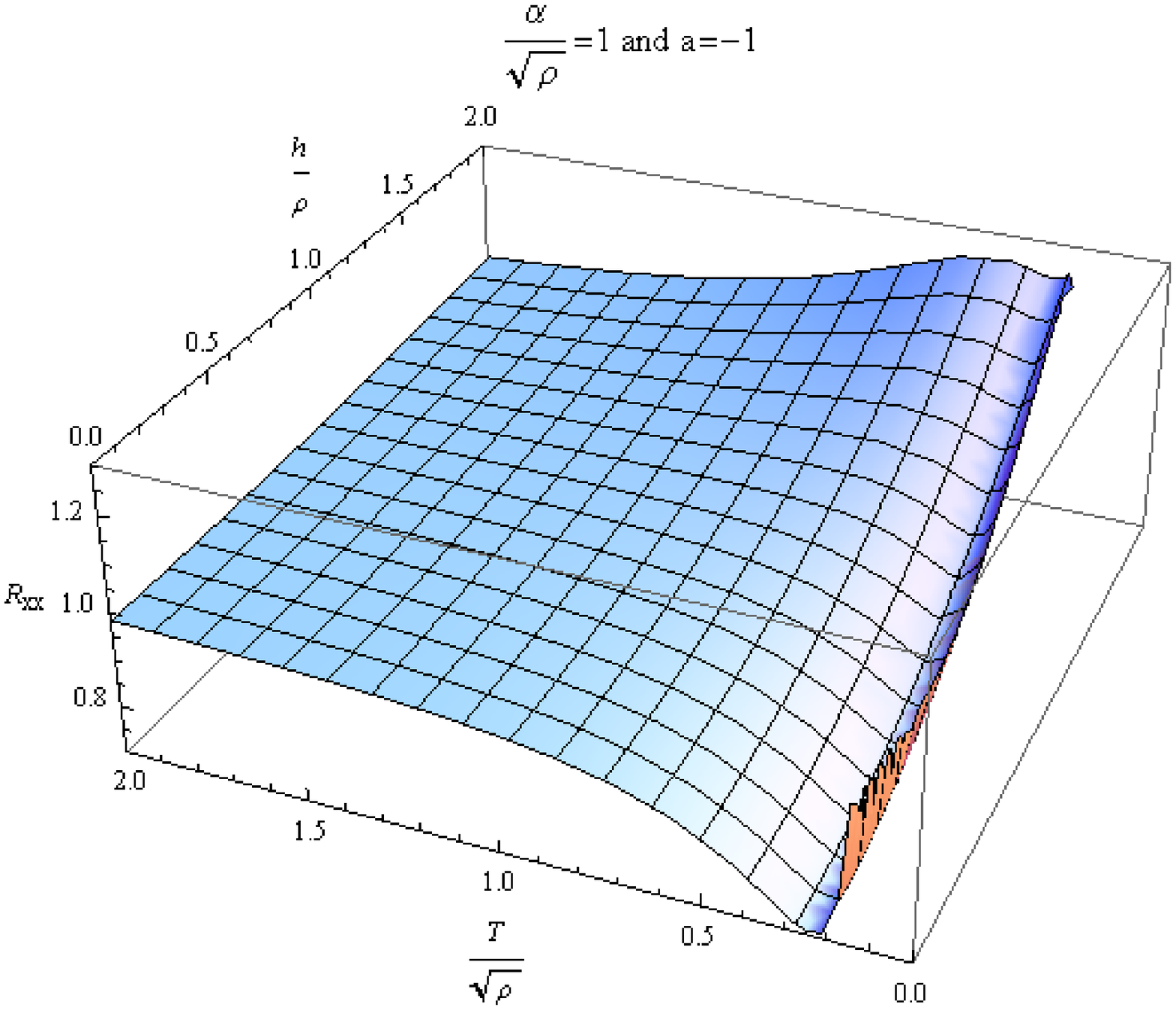}\label{fig:BIMTs3D.eps}}
\subfigure[{Plot of
$R_{xx}$ against $T/\sqrt{\rho}$ for various values of $h/\rho$ for $\alpha/\sqrt{\rho}=1$.}]{
\includegraphics[width=0.45\textwidth]{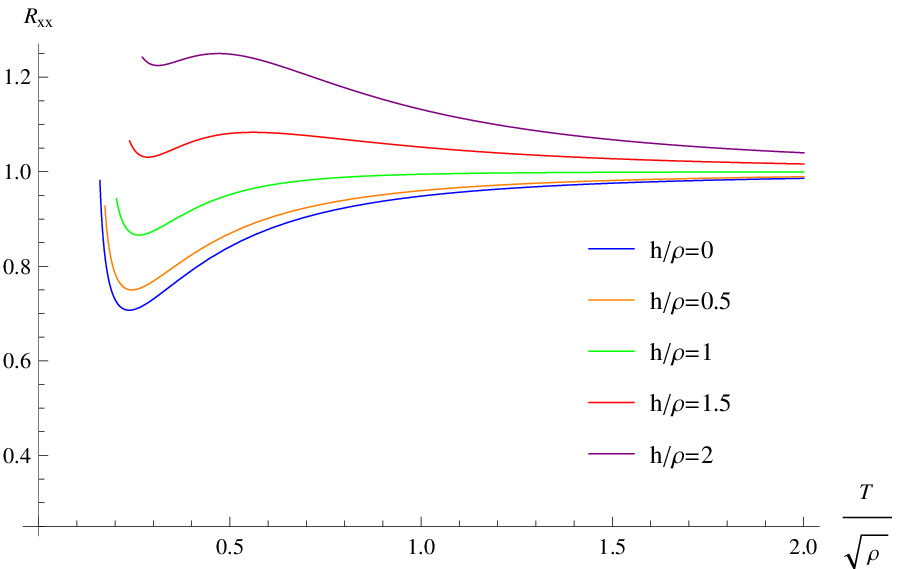}\label{fig:BIMTsT2D}}
\subfigure[{Plot of $R_{xx}$ against $h/\rho$ and $T/\sqrt{\rho}$ for $\alpha/\sqrt{\rho}=10$.}]{
\includegraphics[width=0.35\textwidth]{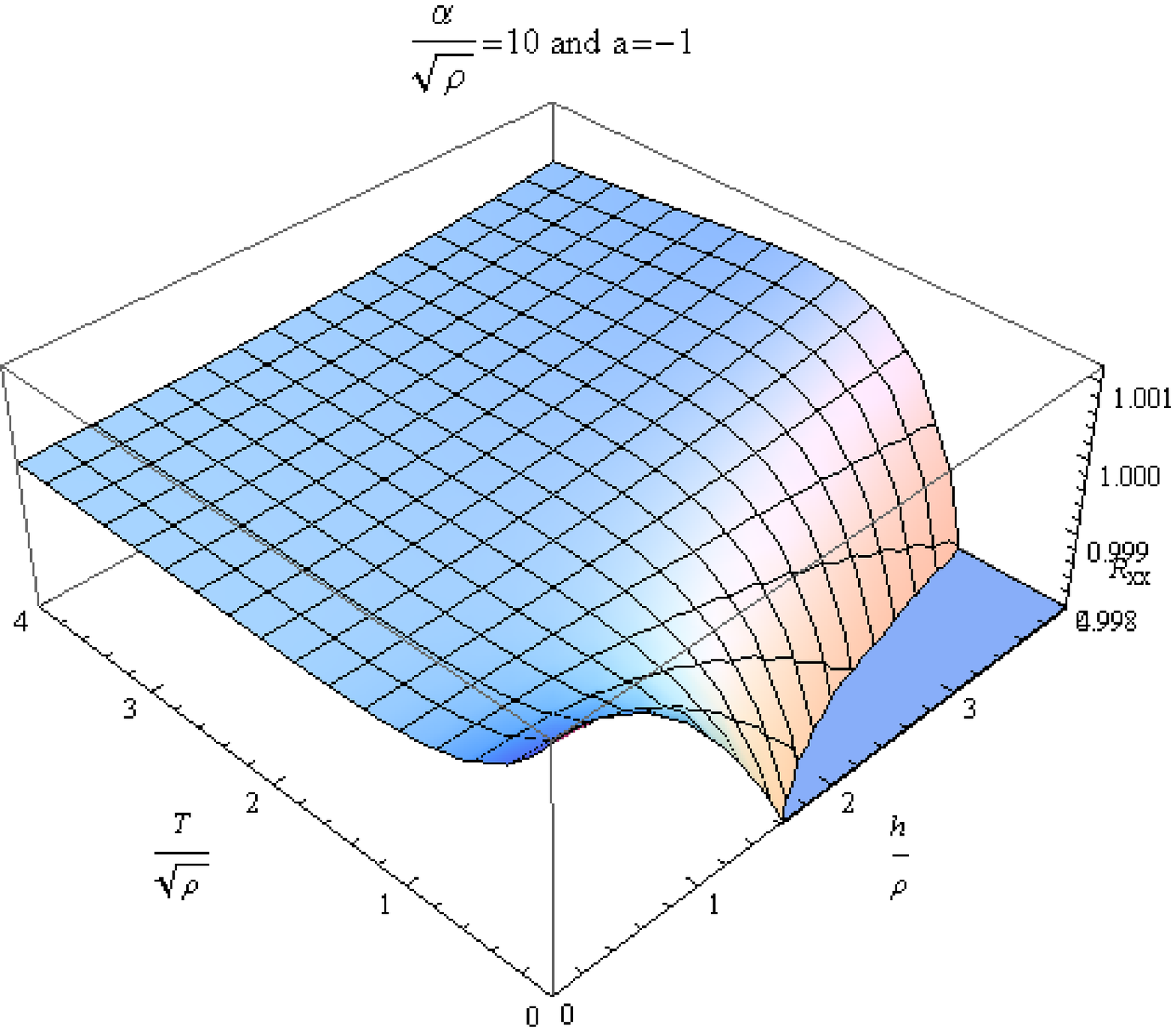}\label{fig:BIMTl3D}}
\subfigure[{Plot of
$R_{xx}$ against $T/\sqrt{\rho}$ for various values of $h/\rho$ for $\alpha/\sqrt{\rho}=10$.}]{
\includegraphics[width=0.45\textwidth]{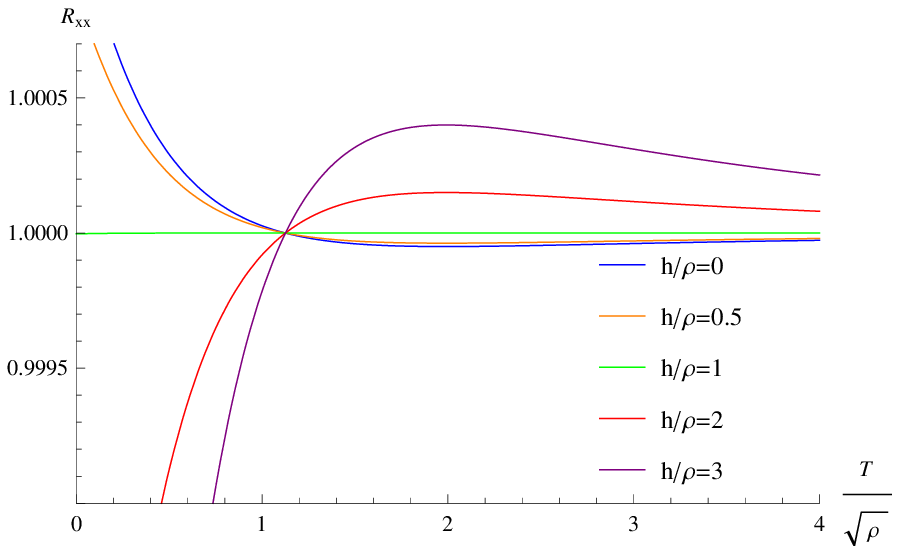}\label{fig:BIMTlT2D}}
\end{center}
\caption{Plots of the temperature dependence of $R_{xx}$ with $\alpha
/\sqrt{\rho}=1$ and $\alpha/\sqrt{\rho}=10$ at finite charge density for
Born-Infeld electrodynamics with $a=-1$.}%
\label{fig:BIMIT}%
\end{figure}

We first discuss behavior of $R_{xx}$ at $T=0$. At zero temperature, the
condition $r_{s}<r_{h}$ gives that there is an upper bound on $h^{2}+\rho^{2}$
for $a<-\frac{1}{6}$:
\begin{equation}
\frac{h^{2}+\rho^{2}}{\alpha^{4}}\leq\frac{-a}{\left(  1+6a\right)  ^{2}}.
\label{eq:upperbound}%
\end{equation}
We plot $R_{xx}$ versus $\rho/\alpha^{2}$ and $h^{2}/\alpha^{2}$ at $T=0$ with
$a=-0.4$ and $a=-1$ in FIGs. \ref{fig:BIMs3D} and \ref{fig:BIMl3D},
respectively. It is noteworthy that, in FIGs. \ref{fig:BIMs3D} and
\ref{fig:BIMl3D}, the domains of $R_{xx}$ are bounded by eqn. $\left(
\ref{eq:upperbound}\right)  $. In FIGs. \ref{fig:BIMsh2D} and
\ref{fig:BIMlh2D}, we display how the resistance $R_{xx}$ depends
$h/\alpha^{2}$ for various values of $\rho/\alpha^{2}$ in the $a=-0.4$ and
$a=-1$ cases, respectively. In both cases, $R_{xx}$ decreases monotonically
with increasing the magnitude of the magnetic field at constant charge
density, which shows that the system exhibits negative magneto-resistance in
all of the allowed parameter range. The resistance $R_{xx}$ as a function of
$\rho/\alpha^{2}$ for different values of $h/\alpha^{2}$ in the $a=-0.4$ case
is presented in FIG. \ref{fig:BIMsrho2D}. We find that $R_{xx}$ increases
monotonically as one increases the magnitude of the charge density with the
magnetic field fixed, which shows that Mott-like behavior occurs in all of the
allowed parameter range. We also display the resistance $R_{xx}$ for $a=-1$ in
FIG. \ref{fig:BIMlrho2D}. When $h=0$, $R_{xx}$ increases monotonically with
increasing the magnitude of the charge density. For a small but non-vanishing
$h/\alpha^{2}$, e.g. $h/\alpha^{2}=0.05$ and $0.1$, the non-monotonic behavior
at large values of $\rho/\alpha^{2}$ appears. As the value of $\rho/\alpha
^{2}$ increases, $R_{xx}$ increases first and then decreases after reaching a
maximum. However for a larger value of $h/\alpha^{2}$, e.g. $h/\alpha
^{2}=0.15$ and $0.19$, we find that $R_{xx}$ decreases monotonically with
increasing the magnitude of the charge density, and hence Mott-like behavior
disappears. In summary, Mott-like behavior always occurs in the $a=-0.4$ case.
However in the $a=-1$ case, Mott-like behavior appears for a weak magnetic
field, and a strong enough magnetic field could destroy it.

\begin{figure}[tb]
\begin{center}
\subfigure[{Plot of $R_{xx}$ against $h/\rho$ at finite charge density with $\alpha
/\sqrt{\rho}=10$ for various values of $T/\sqrt{\rho}$.}]{
\includegraphics[width=0.48\textwidth]{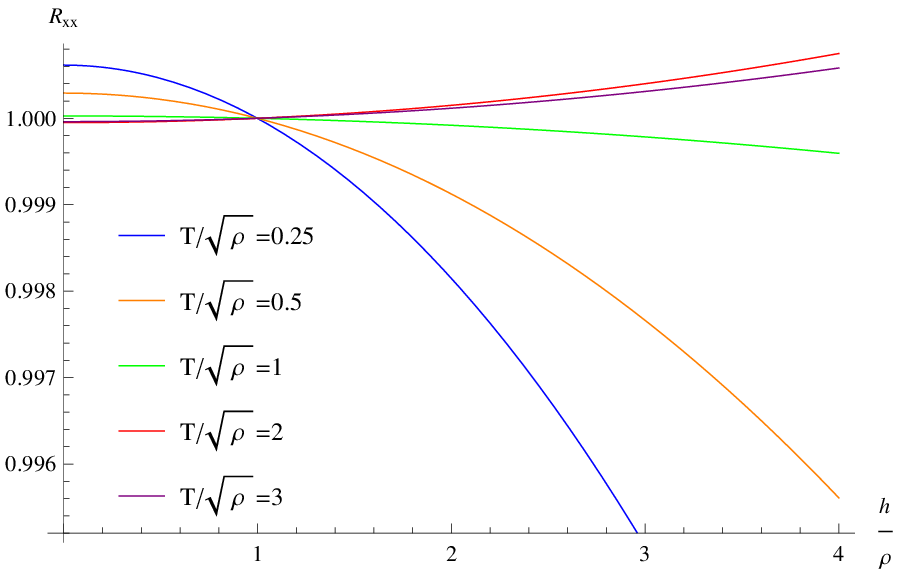}\label{fig:BIMTlh2D}}
\subfigure[{Plot of $R_{xx}$ against $\rho/h$ at finite magnetic field with $\alpha
/\sqrt{h}=10$ for various values of $T/\sqrt{h}$.}]{
\includegraphics[width=0.48\textwidth]{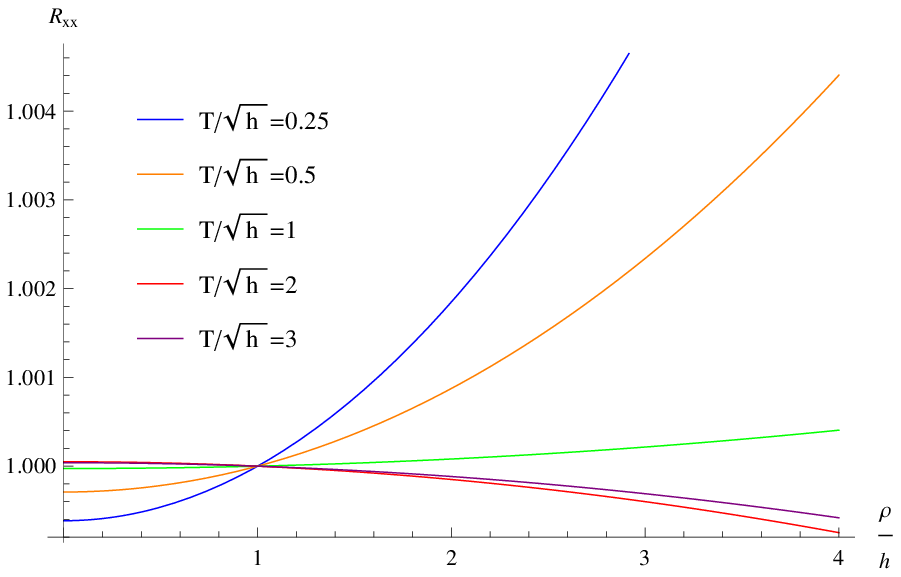}\label{fig:BIMTlrho2D}}
\end{center}
\caption{Plots of how the dependences of $R_{xx}$ on $\rho$ and $h$ change
with respect to $T$. Here we consider Born-Infeld electrodynamics with
$a=-1$.}%
\end{figure}

Next, we consider the temperature dependence of $R_{xx}$ at finite charge
density. Focusing on the $a=-1$ case, we present how $R_{xx}$ depends on
$h/\rho$ and $T/\sqrt{\rho}$ for $\alpha/\sqrt{\rho}=1$ and $\alpha/\sqrt
{\rho}=10$ in FIG. \ref{fig:BIMIT}. At zero temperature, eqn. $\left(
\ref{eq:upperbound}\right)  $ would put an upper bound on the value of
$h/\rho$
\begin{equation}
\left(  h/\rho\right)  ^{2}\leq\frac{\left(  \alpha/\sqrt{\rho}\right)  ^{4}%
}{25}-1.
\end{equation}
When $\alpha/\sqrt{\rho}=1$, the RHS of the above equation is negative, which
explains why the curves in FIG. \ref{fig:BIMTsT2D} could not go to zero
temperature. Since Born-Infeld electrodynamics is CP invariant, one has
insulating behavior for $h/\rho>1$ and metallic behavior for $h/\rho<1$ in the
high temperature limit, which is clearly shown in FIGs. \ref{fig:BIMTsT2D} and
\ref{fig:BIMTlT2D}. However, at low temperatures, the temperature dependence
of $R_{xx}$ in the $a<0$ case is quite different from those in the $a>0$ and
Maxwell cases. For $\alpha/\sqrt{\rho}=1$, FIG. \ref{fig:BIMTsT2D} displays
insulating behavior at low temperatures. As one increases the temperature, the
system would start to exhibit metallic behavior. If one keeps increasing the
temperature, the system would stay metallic behavior for $h/\rho<1$, but it
would return to insulating behavior for $h/\rho<1$. In the $\alpha/\sqrt{\rho
}=10$ case, according to FIG. \ref{fig:BIMTlT2D}, one has insulating behavior
for $h/\rho<1$ and metallic behavior for $h/\rho>1$ at low temperatures.
Therefore, increasing the magnitude of the magnetic field would induce a
transition or crossover from insulating to metallic behavior at low
temperatures and that from metallic to insulating behavior at high temperatures.

We find that the system does not exhibit negative magneto-resistance or
Mott-like behavior at high temperatures. If one has negative
magneto-resistance or Mott-like behavior at low temperatures, they would
disappear at a high enough temperature. We plot $R_{xx}$ as a function of
$h/\rho$ for various values of $T/\sqrt{\rho}$ with $\alpha/\sqrt{\rho}=10$ in
FIG. \ref{fig:BIMTlh2D}. With a fixed value of the charge density, FIG.
\ref{fig:BIMTlh2D} shows that $R_{xx}$ decreases with increasing $h$ for
$T/\sqrt{\rho}=0.25$, $0.5$ and $1$, and $R_{xx}$ increases with increasing
$h$ for $T/\sqrt{\rho}=2$ and $3$. Similarly, FIG. \ref{fig:BIMTlrho2D} shows
that, with a fixed value of the magnetic field, the system displays Mott-like
behavior for $T/\sqrt{h}=0.25$, $0.5$ and $1$, and $R_{xx}$ decreases with
increasing $\rho$ for $T/\sqrt{h}=2$ and $3$.

\subsection{Square Electrodynamics}

\begin{figure}[tb]
\begin{center}
\subfigure[{Plot of $R_{xx}$ against $h/\alpha^{2}$ and $\rho/\alpha^{2}$ at $T=0$ with $a=-0.4$.}]{
\includegraphics[width=0.35\textwidth]{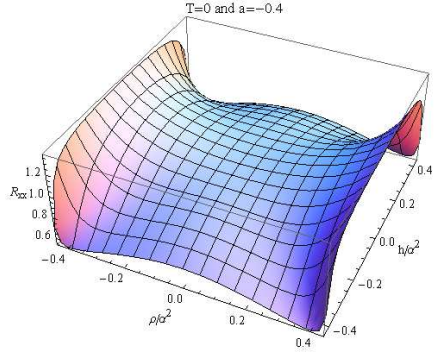}\label{fig:SFs3D}}
\subfigure[{Plot of $R_{xx}$ against $\rho/\alpha^{2}$ for various values of $h
/\alpha^{2}$ at $T=0$ with $a=-0.4$.}]{
\includegraphics[width=0.4\textwidth]{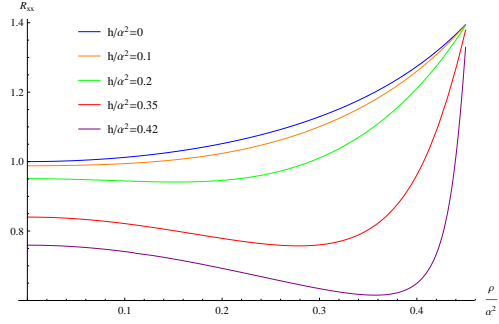}\label{fig:SFs2D}}
\subfigure[{Plot of $R_{xx}$ against $h/\alpha^{2}$ and $\rho/\alpha^{2}$ at $T=0$ with $a=-1$.}]{
\includegraphics[width=0.35\textwidth]{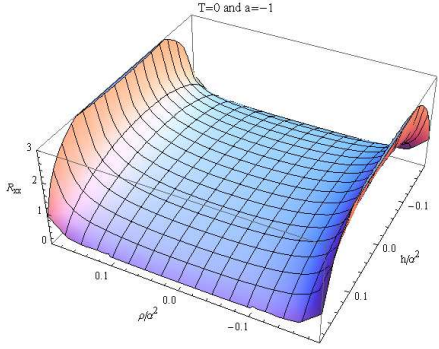}\label{fig:SFl3D}}
\subfigure[{Plot of $R_{xx}$ against $\rho/\alpha^{2}$ for various values of $h
/\alpha^{2}$ at $T=0$ with $a=-1$.}]{
\includegraphics[width=0.4\textwidth]{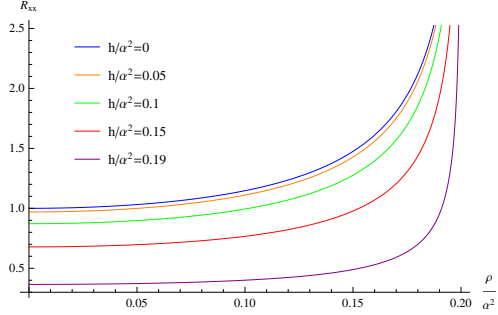}\label{fig:SFl2D}}
\end{center}
\caption{Plots of $R_{xx}$ at $T=0$ for square electrodynamics. Top figures:
$a=-0.4$. Bottom figures: $a=-1$. }%
\end{figure}

Consider a Born--Infeld like Lagrangian
\begin{equation}
\mathcal{L}\left(  s,p\right)  =\frac{1}{a}\left(  1-\sqrt{1-2as}\right)
\text{,}%
\end{equation}
which gives the same result for $\sigma_{ij}/R_{ij}$ as Born-Infeld
electrodynamics in the $h=0$ case. We now study the dependence of $R_{xx}$ on
$\rho$ and $h$ at $T=0$. Since the behavior of $R_{xx}$ in the $a>0$ case is
similar to that in Maxwell electrodynamics, we focus on the $a<0$ case. In
FIG. \ref{fig:SFs3D}, we plot $R_{xx}$ as a function of $\rho/\alpha^{2}$ and
$h^{2}/\alpha^{2}$ with $a=-0.4$, which displays negative magneto-resistance
at fixed charge density for all of the allowed parameter range. FIG.
\ref{fig:SFs2D} shows the dependence of $R_{xx}$ on $\rho/\alpha^{2}$ for
various values of $h^{2}/\alpha^{2}$. For a small value of $h/\alpha^{2}$,
e.g. $h/\alpha^{2}=0$ and $0.1$, $R_{xx}$ increases monotonically with
increasing $\rho/\alpha^{2}$. However for a larger value of $h/\alpha^{2}$,
e.g. $h/\alpha^{2}=0.2$, $0.35$ and $0.42$, $R_{xx}$ first decreases, then
reaches a minimum, and then increases monotonically with increasing
$\rho/\alpha^{2}$. In this case, Mott-like behavior would appear for large
enough values of $\rho/\alpha^{2}$. On the other hand, according to FIGs.
\ref{fig:SFl3D} and \ref{fig:SFl2D}, the system exhibits negative
magneto-resistance and Mott-like behavior for all of the allowed parameter
range in the $a=-1$ case.

\subsection{Logarithmic Electrodynamics}

\begin{figure}[tb]
\begin{center}
\subfigure[{Plot of $R_{xx}$ against $h/\alpha^{2}$ and $\rho/\alpha^{2}$ at $T=0$ with $a=-1$.}]{
\includegraphics[width=0.35\textwidth]{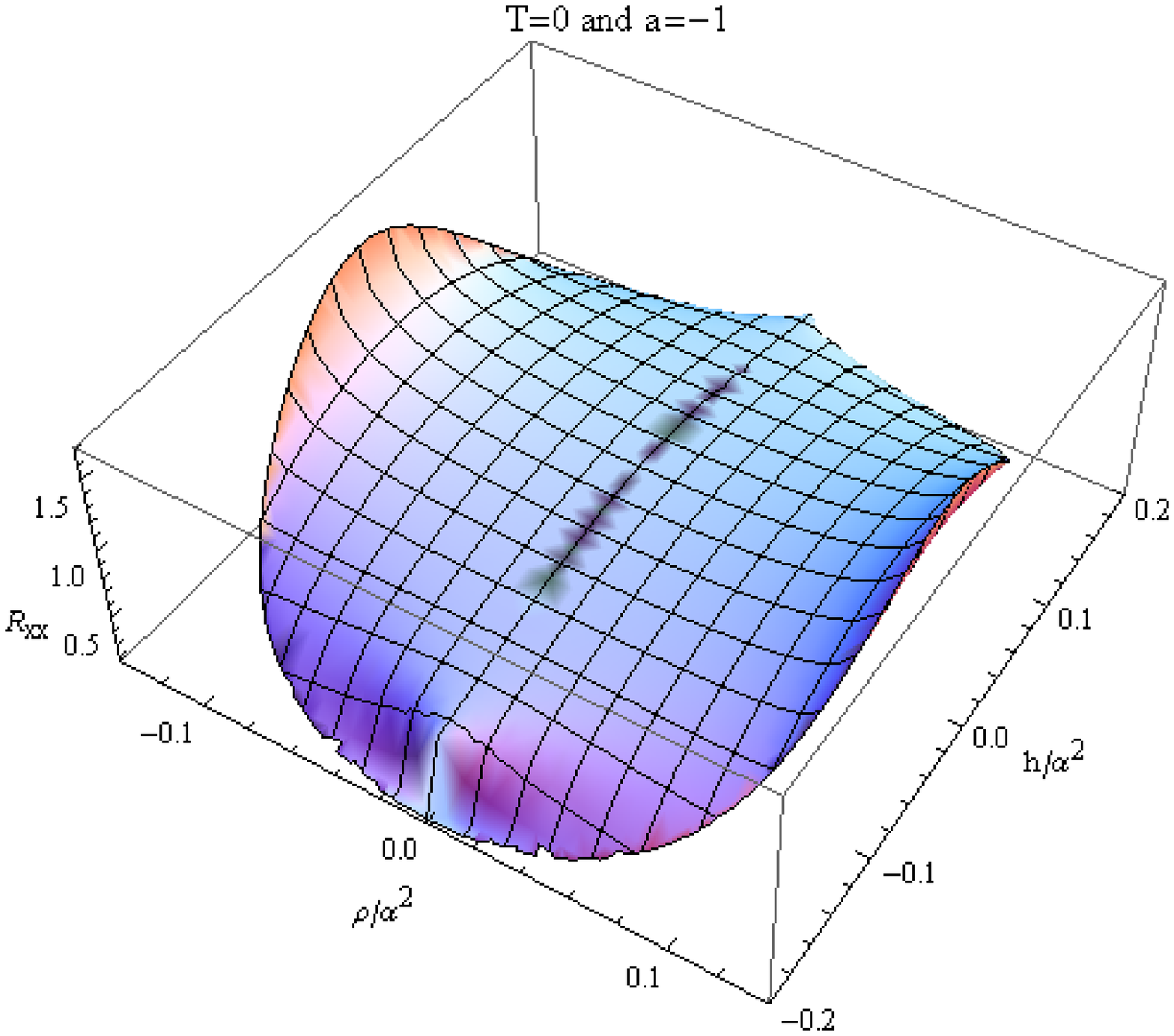}\label{fig:Log3D}}
\subfigure[{Plot of $R_{xx}$ against $\rho/\alpha^{2}$ for various values of $h
/\alpha^{2}$ at $T=0$ with $a=-1$.}]{
\includegraphics[width=0.4\textwidth]{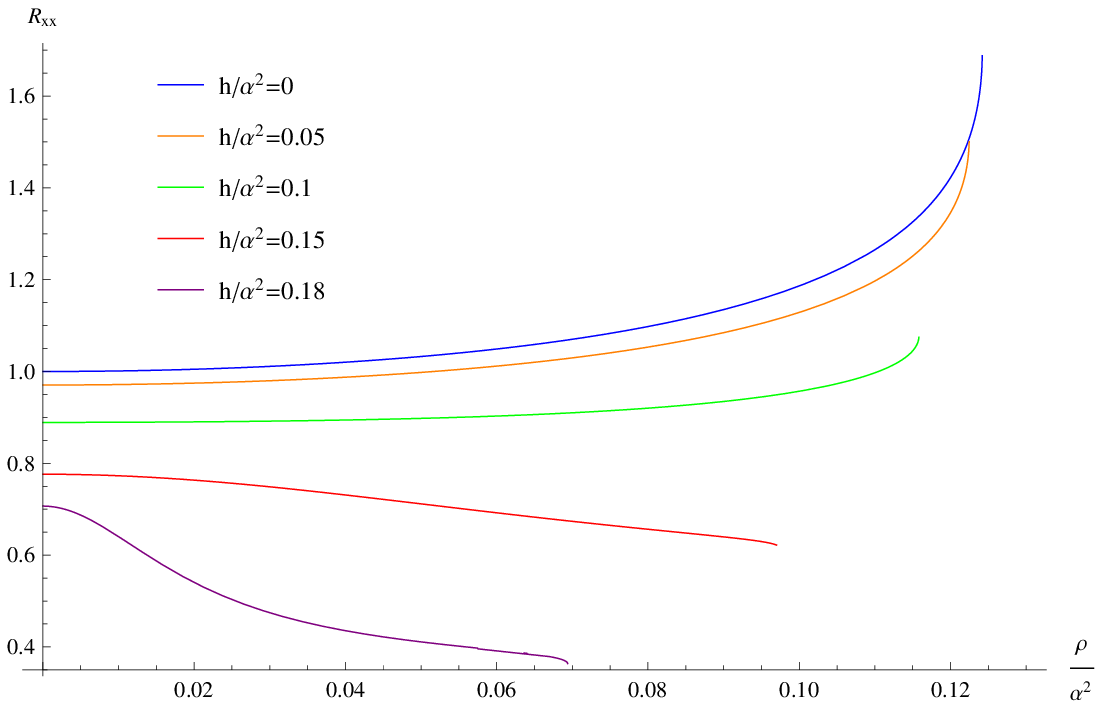}\label{fig:Log2D}}
\end{center}
\caption{Plots of $R_{xx}$ at $T=0$ for logarithmic electrodynamics with
$a=-1$. }%
\label{fig:Log}%
\end{figure}

Finally, we consider logarithmic electrodynamics, whose Lagrangian is
described by%
\begin{equation}
\mathcal{L}\left(  s,p\right)  =-\frac{1}{a}\log\left(  1-as-\frac{a^{2}p^{2}%
}{2}\right)  .
\end{equation}
We display the dependence of $R_{xx}$ on $\rho/\alpha^{2}$ and $h/\alpha^{2}$
in the the $a=-1$ case at $T=0$ in FIG. \ref{fig:Log3D}, which shows that
$\partial R_{xx}/\partial\left\vert h\right\vert <0$. However, FIG
\ref{fig:Log2D} shows that $\partial R_{xx}/\partial\left\vert \rho\right\vert
>0$ for small values of $h/\alpha^{2}$, and $\partial R_{xx}/\partial
\left\vert \rho\right\vert <0$ for large enough values of $h/\alpha^{2}$,
which means a strong enough magnetic field would destroy Mott-like behavior.

\section{Discussion and Conclusion}

\label{Sec:Con}

\begin{table}[!htbp]
\centering
\begin{tabular}
[c]{|m{0.7in}<{\centering}|m{1.1in}<{\centering}|m{0.7in}<{\centering}|m{1.85in}<{\centering}|m{1.8in}<{\centering}|}\hline
& Lagrangian & Parameter & $\rho$ dependence of $R_{xx}$ & $h$ dependence of
$R_{xx}$\\\hline
\multicolumn{1}{|m{0.7in}<{\centering}|}{{\footnotesize Maxwell}} & $s$ &  &
{\footnotesize $\partial_{\left\vert \rho\right\vert }R_{xx}<0$. See FIG.
\ref{fig:MaxwellT=0}.} & {\footnotesize $\partial_{\left\vert h\right\vert
}R_{xx}>0$. See FIG. \ref{fig:MaxwellT=0}.}\\\hline
\multicolumn{1}{|m{0.7in}<{\centering}|}{{\footnotesize Maxwell-Chern-Simons}} & $s+\theta
p$ & $\theta\neq0$ & {\footnotesize There exists parameter space region for
$\partial_{\left\vert \rho\right\vert }R_{xx}>0$. }{\footnotesize See FIG.
\ref{fig:MOTTCS}. } & {\footnotesize There exists parameter space region for
$\partial_{\left\vert h\right\vert }R_{xx}<0$. }{\footnotesize See eqn.
$\left(  \ref{eq:csnm}\right)  .$ }\\\hline
\multicolumn{1}{|m{0.7in}<{\centering}|}{} &  & $a>0$ & {\footnotesize $\partial
_{\left\vert \rho\right\vert }R_{xx}<0$. See FIG. \ref{fig:BIP3D}. } &
{\footnotesize $\partial_{\left\vert h\right\vert }R_{xx}>0$. See FIG.
\ref{fig:BIP3D}.}\\\cline{3-5}%
{\footnotesize Born-Infeld} & $\frac{1-\sqrt{1-2as-a^{2}p^{2}}}{a}$ & $a=-0.4$
& {\footnotesize $\partial_{\left\vert \rho\right\vert }R_{xx}>0$. See FIG.
\ref{fig:BIMsrho2D}. } & {\footnotesize $\partial_{\left\vert h\right\vert
}R_{xx}<0$. See FIG. \ref{fig:BIMsh2D}. }\\\cline{3-5}%
\multicolumn{1}{|m{0.7in}<{\centering}|}{} &  & $a=-1$ & {\footnotesize $\partial
_{\left\vert \rho\right\vert }R_{xx}>0$\ for small values of $h/\alpha^{2}$
and $\rho/\alpha^{2}$. See FIG. \ref{fig:BIMlrho2D}.} &
{\footnotesize $\partial_{\left\vert h\right\vert }R_{xx}<0$. See FIG.
\ref{fig:BIMlh2D}.}\\\hline
\multicolumn{1}{|m{0.7in}<{\centering}|}{} &  & $a>0$ & {\footnotesize $\partial
_{\left\vert \rho\right\vert }R_{xx}<0$. } & {\footnotesize $\partial
_{\left\vert h\right\vert }R_{xx}>0.$}\\\cline{3-5}%
{\footnotesize Square } & $\frac{1-\sqrt{1-2as}}{a}$ & $a=-0.4$ &
{\footnotesize $\partial_{\left\vert \rho\right\vert }R_{xx}>0$\ for small
values of }$h/\alpha^{2}${\footnotesize . For larger values of $h/\alpha^{2}$,
$\partial_{\left\vert \rho\right\vert }R_{xx}>0$} {\footnotesize only for
large enough values of $\rho/\alpha^{2}$. See FIG. \ref{fig:SFs2D}.} &
{\footnotesize $\partial_{\left\vert h\right\vert }R_{xx}<0$. See FIG.
\ref{fig:SFs3D}. }\\\cline{3-5}%
\multicolumn{1}{|m{0.7in}|}{} &  & $a=-1$ & {\footnotesize $\partial
_{\left\vert \rho\right\vert }R_{xx}>0$. \ See FIG. \ref{fig:SFl2D}.} &
{\footnotesize $\partial_{\left\vert h\right\vert }R_{xx}<0$. See FIG.
\ref{fig:SFl3D}.}\\\hline
\multicolumn{1}{|m{0.7in}|}{} &  & $a>0$ & {\footnotesize $\partial
_{\left\vert \rho\right\vert }R_{xx}<0.$} & {\footnotesize $\partial
_{\left\vert h\right\vert }R_{xx}>0.$}\\\cline{3-5}%
{\footnotesize Logarithmic} & $\frac{-\log\left(  1-as-\frac{a^{2}p^{2}}%
{2}\right)  }{a}$ & $a=-1$ & {\footnotesize $\partial_{\left\vert
\rho\right\vert }R_{xx}>0$\ for small values of $h/\alpha^{2}$. See FIG.
\ref{fig:Log2D}. } & {\footnotesize $\partial_{\left\vert h\right\vert }%
R_{xx}<0$. See FIG. \ref{fig:Log3D}.}\\\hline
\end{tabular}
\caption{The dependence of the in-plane resistance $R_{xx}$ on $\rho/\alpha^{2}$ and
$h/\alpha^{2}$ at $T=0$. Here $\alpha$ is a parameter responsible for
generating momentum dissipation. Note that $\partial_{\left\vert h\right\vert
}R_{xx}<0$ means negative magneto-resistance while $\partial_{\left\vert
\rho\right\vert }R_{xx}<0$ means Mott-like behavior.}
\label{tab:T=0}
\end{table}

In this paper, we used gauge/gravity duality to investigate the properties of
the DC conductivities with a finite magnetic field of a strongly correlated
system in $2+1$ dimensions. The charge current in the boundary field theory is
dual to a NLED field in bulk. In our holographic setup, we considered the
backreaction effects of the NLED field on the geometry and introduced axionic
scalars to generate momentum dissipation. We then presented the expressions
for the DC conductivities for a general NLED field. Specifically, one can use
eqns. $\left(  \ref{eq:HT}\right)  $, $\left(  \ref{eq:rho}\right)  $ and
$\left(  \ref{eq:DCconductity}\right)  $ to express the DC conductivities in
terms of the temperature $T$, the charge density $\rho$ and the magnetic field
$h$ of the dual field theory. In the second part of our paper, we discussed
the properties of the in-plane resistance $R_{xx}$\ in some interesting NLED
models, where there appeared Mott-like behavior or negative magneto-resistance
in some cases. In Table \ref{tab:T=0}, we summarize the results for the $\rho$
and $h$ dependences of $R_{xx}$\ at zero temperature in the NLED models
discussed above.

Table \ref{tab:T=0} shows that the behavior of $R_{xx}$ as a function of
$\rho$ and $h$ is sensitive to the sign of the parameter $a$. To shed light on
the role of $a$, we calculate the correction to the coulomb force between two
electrons due to non-linearities from the NLED Lagrangian $\mathcal{L}\left(
s,p\right)  $. The corrected coulomb force is given by
\begin{equation}
F\approx\frac{e^{2}}{4\pi r^{2}}\left[  1-\frac{\mathcal{L}^{\left(
2,0\right)  }\left(  0,0\right)  }{2}\left(  \frac{e}{4\pi r^{2}}\right)
^{2}\right]  .
\end{equation}
For the NLED models discussed in section \ref{Sec:Examples}, we have
$\mathcal{L}^{\left(  2,0\right)  }\left(  0,0\right)  =a$. For $a>0$, the
non-linearities correction tends to reduce the strength of the repulsive force
between two electrons. However for $a<0$, the correction tends to increase the
strength of the force. So it is natural to expect that a negative $a$ may
correspond to strong interactions between electrons in our holographic model,
which could lead to Mott-like behavior $^{\left[  \ref{ft:2}\right]  }%
$\footnotetext[1]{\label{ft:2} In \cite{IN-Baggioli:2016oju}, the role of $a$
has been discussed for iDBI model.}. \

It is interesting to note that Born-Infeld electrodynamics with $a=-1$ could
describe Mott insulator to metal transition (IMT) induced by a magnetic field.
In fact, FIG. \ref{fig:BIMTlT2D} shows that, at low temperatures, the system
has insulating/metallic behavior for a weak/strong magnetic field. On the
other hand, FIG. \ref{fig:BIMlrho2D} displays that, for a weak magnetic field
at low temperatures, the system usually has Mott-like behavior and hence is a
Mott-Insulator. When the magnetic field grows strong enough, Mott-like
behavior disappears, and meanwhile, the system exhibits metallic behavior. A
magnetic field-induced IMT for a Mott system, namely a bilayer ruthenate,
Ti-doped Ca$_{3}$Ru$_{2}$O$_{7}$, was presented in \cite{CON-Zhu2016}. Our
analysis for IMT is rather qualitative, and it deserves future more detailed studies.

For a negative enough $a$, we found that, at low temperatures, the resistance
$R_{xx}$ always decreases with increasing magnetic field, which appears as
negative magneto-resistance. On the other hand, the behavior of $R_{xx}$ in
the $a>0$ NLED models is similar to that in Maxwell electrodynamics, in which
one always have positive-resistance. It seems that, decreasing $a$ from a
positive value to a negative one, which corresponds to increasing the strength
of the interactions between electrons, would lead the magneto-resistance to
change from positive to negative. It showed in \cite{CON-Zhou2011} that the
magneto-resistance could change from positive to negative by gradually
introducing artificial disorder through Ga$^{+}$ ion irradiation to pristine
graphene. To relate our results to the experiments, we need to better
understand how $a$ is related to external control parameters.

In this paper, we found the $\sigma_{ij}/R_{ij}$ expressions for a general
NLED field. Our analyses for the properties of the resistance in NLED models
are preliminary. One can use these expressions to find or construct a NLED
model to realize some interesting experimental results, such as the scaling
relationship between applied magnetic field and temperature observed in the
magneto-resistance of the pnictide superconductor.

\begin{acknowledgments}
We are grateful to Zheng Sun for useful discussions and valuable comments.
This work is supported in part by NSFC (Grant No. 11005016, 11175039 and 11375121).
\end{acknowledgments}

\end{document}